\documentclass{aa}

\makeatletter 
\renewcommand*{\aa@journalname}{} 
\renewcommand*{\aa@manuscriptname}{} 
\makeatother

\usepackage{graphicx}
\usepackage[varg]{txfonts}
\usepackage[colorlinks=true, allcolors=blue]{hyperref}
\usepackage{mathrsfs}
\usepackage{tabularx}
\usepackage{booktabs}
\usepackage[autostyle, english = american]{csquotes}

\usepackage{newunicodechar}
\newunicodechar{Λ}{\ensuremath{\Lambda}}
\newunicodechar{γ}{\ensuremath{\gamma}}
\newunicodechar{₀}{\ensuremath{{}_0}}
\newunicodechar{Ω}{\ensuremath{\Omega}}
\newunicodechar{ν}{\ensuremath{\nu}}
\newunicodechar{τ}{\ensuremath{\tau}}
\newunicodechar{ℋ}{\ensuremath{\mathscr{H}}}
\newunicodechar{Φ}{\ensuremath{\Phi}}
\newunicodechar{Ψ}{\ensuremath{\Psi}}
\newunicodechar{ₐ}{\ensuremath{{}_a}}
\newunicodechar{ₛ}{\ensuremath{{}_s}}
\newunicodechar{ρ}{\ensuremath{\rho}}
\newunicodechar{δ}{\ensuremath{\delta}}
\newunicodechar{θ}{\ensuremath{\theta}}
\newunicodechar{σ}{\ensuremath{\sigma}}
\newunicodechar{π}{\ensuremath{\pi}}

\usepackage{fancyvrb}
\usepackage{mdframed}
\newmdenv[
  backgroundcolor=black!5,
  linewidth=0pt,
  roundcorner=0pt,
  innerlinewidth=0pt,
  innertopmargin=5pt,
  innerbottommargin=5pt,
  innerleftmargin=5pt,
  innerrightmargin=5pt,
  font=\small,
]{codebox}

\usepackage{tikz}
\usetikzlibrary{arrows}
\usetikzlibrary{positioning}
\usetikzlibrary{arrows.meta}
\usetikzlibrary{trees}

\DeclareMathOperator{\sinc}{sinc}
\newcommand{\LCDM}{$\mathrm{\Lambda C D M}$}
\newcommand\scrH{\mathscr{H}}
\newcommand\lmax{l_\text{max}}
\newcommand\dlnfdlnx{\frac{\mathrm{d} \ln f}{\mathrm{d} \ln x}}
\newcommand\el{\mathrm{e}}
\newcommand\Hy{\mathrm{H}}
\newcommand\He{\mathrm{He}}
\newcommand\Hesin{{\mathrm{He}_1}}
\newcommand\Hetri{{\mathrm{He}_3}}
\newcommand\reone{{\mathrm{re}_1}}
\newcommand\retwo{{\mathrm{re}_2}}

\begin{document}

\title{SymBoltz.jl: A symbolic-numeric, approximation-free, and differentiable linear Einstein–Boltzmann solver}
\authorrunning{Sletmoen, H.}
\titlerunning{SymBoltz.jl: A symbolic-numeric, approximation-free, and differentiable linear Einstein–Boltzmann solver}
\author{Herman Sletmoen\thanks{Email: \texttt{herman.sletmoen@astro.uio.no}}}
\institute{Institute of Theoretical Astrophysics, University of Oslo, PO Box 1029 Blindern, 0315 Oslo, Norway}
\date{Received XX / Accepted XX}
\abstract{
SymBoltz is a new Julia package for solving the linear Einstein–Boltzmann equations in cosmology.
It features a symbolic-numeric interface for specifying equations, is free of approximation switching schemes, and is compatible with automatic differentiation.
Cosmological models are built from replaceable physical components in a way that scales well to extended models, or alternatively written as one compact system of equations.
The modeler provides their equations, and SymBoltz solves them while reducing friction in the modeling process.
Symbolic knowledge enables powerful automation of tasks, such as separating computational stages (e.g., background and perturbations), generating analytical and sparse Jacobian matrices, and interpolating arbitrary variables from the solution.
Implicit solvers integrate the full stiff equations at all times without approximations, which greatly simplifies the code.
Performance remains comparable to existing approximation-based codes due to efficient high-order implicit methods, fast generated code, optimal handling of the Jacobian, and sparse matrix methods.
Automatic differentiation gives exact derivatives of any output with respect to any input, which is important for gradient-based Markov chain Monte Carlo (MCMC) sampling in large parameter spaces, Fisher forecasting, emulator training, and sensitivity analyses.
The main features form a synergy that reinforces the design of the code.
Output spectra agree with established codes up to 0.1\% with standard precision levels.
More work is needed to implement additional features and for fast reverse-mode automatic differentiation of scalar loss functions.
SymBoltz is publicly available with single-command installation and extensive documentation.
We welcome all contributions to the code from the community.
}


\maketitle

\nolinenumbers

\section{Introduction}
\label{sec:intro}

Cosmology is at a crossroads \citep{freedmanCosmologyCrossroads2017}, as the precision of modern observations continues to reveal tensions with theory.
This suggests that the $\Lambda$ cold dark matter (\LCDM{}) model is incomplete, despite successfully explaining many observations.
On the theoretical side, this situation drives a search for more realistic models through modifications to \LCDM{} \citep{bull$L$CDMProblemsSolutions2016}.
This exploration benefits from access to numerical tools that are easy to modify, encouraging us to relax model-dependent approximations, create user-friendly interfaces, and structure codes with modular components.
Linear Einstein–Boltzmann solvers (\enquote{Boltzmann codes}), such as CAMB \citep{lewisEfficientComputationCMB2000a} and CLASS \citep{lesgourguesCosmicLinearAnisotropy2011}, are essential tools in the cosmological modeling toolbox.

On the observational frontier, next-generation surveys such as the
Square Kilometre Array \citep{dewdneySquareKilometreArray2009},
\textit{Vera C. Rubin} Observatory \citep{lsstsciencecollaborationLSSTScienceBook2009},
Dark Energy Spectroscopic Instrument \citep{desicollaborationDESIExperimentPart2016},
Simons Observatory \citep{thesimonsobservatorycollaborationSimonsObservatoryScience2019}, and
\textit{Euclid} \citep{euclidcollaborationEuclidOverviewEuclid2025}
promise more precise data.
Setting models apart with upcoming data involves both theoretical model parameters and experimental nuisance parameters that coexist in large $O(100)$-dimensional spaces \citep{pirasFutureCosmologicalLikelihoodbased2024}.
In these high dimensions, modern Markov chain Monte Carlo (MCMC) methods, such as Hamiltonian Monte Carlo and the No-U-Turn Sampler \citep{hoffmanNoUTurnSamplerAdaptively2014}, outperform the traditional Metropolis–Hastings algorithm \citep{hastingsMonteCarloSampling1970}.
They use both the likelihood and its gradient to explore parameter space faster.
Emulators have become a popular proxy to get gradients from nondifferentiable codes by training differentiable neural networks to reproduce their output \citep[e.g.,][]{bollietHighaccuracyEmulatorsObservables2024,boniciCapsejlEfficientAutodifferentiable2024}.
Differentiability is also used in forward-modeling field-level inference with simulations and initial conditions from Boltzmann solvers \citep{seljakOptimalExtractionCosmological2017}.
Automatic differentiation can compute derivatives more accurately and faster than approximate and brute-force finite differences.
This makes directly differentiable Boltzmann solvers a key addition to the modern cosmological modeling toolkit.

SymBoltz.jl\footnote{\url{https://github.com/hersle/SymBoltz.jl}} is a new Boltzmann solver that aims to meet these needs, featuring a convenient symbolic-numeric interface, approximation-freeness, and differentiability.
At its core, a Boltzmann code solves the Einstein equations for a gravitational theory coupled to some particle species described by Boltzmann equations up to first perturbative order around a homogeneous and isotropic universe.
For example, it predicts cosmic microwave background (CMB), baryon acoustic oscillation (BAO), or supernova (SN) observations, and generates initial conditions for nonlinear $N$-body simulations of large-scale structure.

This article is structured as follows.
Section \ref{sec:intro} motivates SymBoltz by reviewing the history of Boltzmann codes and gradient methods.
Section \ref{sec:features} presents its structure and main features.
Section \ref{sec:examples} shows example use.
Section \ref{sec:performance} compares performance to CLASS.
Section \ref{sec:futurework} outlines paths for future work.
Section \ref{sec:conclusion} concludes with the current state of the code.
Appendices \ref{sec:implementation}, \ref{sec:precision} and \ref{sec:testing} list equations and technical details.

Historically, \citet{peeblesPrimevalAdiabaticPerturbation1970} were the first to numerically integrate a comprehensive set of linear Einstein–Boltzmann equations.
Their work was refined over several years, and \citet{maCosmologicalPerturbationTheory1995} laid the groundwork for modern Boltzmann solvers with the COSMICS code.
Shortly after, \citet{seljakLineSightApproach1996} integrated the photon multipoles by parts to reduce thousands of coupled differential equations to independent integral solutions.
This line-of-sight integration method was first implemented in their CMBFAST code and has become a standard technique that greatly speeds up the calculation, but requires truncating the multipoles in the differential equations.
This Fortran codebase has since evolved into CAMB\footnote{\url{https://camb.info/}} written by \citet{lewisEfficientComputationCMB2000a}, which is one of the two most used and maintained Boltzmann solvers today.
\citet{doranCMBEASYObjectOriented2005} ported CMBFAST to C++ with the fork CMBEASY, which structured the code in an object-oriented fashion and improved user-friendliness, but this project is abandoned today.

\citet{lesgourguesCosmicLinearAnisotropy2011} and \citet{blasCosmicLinearAnisotropy2011a} started the second major family of Boltzmann solvers with the birth of CLASS\footnote{\url{http://class-code.net}} in C.
It improved performance, user-friendliness, code flexibility, ease of modification, and control over precision parameters.
It was the first independent cross-check of CAMB and boosted the scientific accuracy of the Planck mission.

Since then a healthy arms race has fueled refinements to CAMB and CLASS.
Both codes have spawned many forks for studying alternative models and performing custom calculations.
Today they are very well-made, efficient, and reliable tools.

Recently, several alternative solvers have appeared on the market.
They target new numerical techniques, lack of approximations, differentiability, GPU parallelization, interactivity, and symbolic computation.
PyCosmo\footnote{\url{https://pypi.org/project/PyCosmo/}} by \citet{refregierPyCosmoIntegratedCosmological2018} in Python was the first symbolic-numeric Boltzmann code that generates fast C++ code from user-provided symbolic equations, optimizes the sparse and analytical Jacobian matrix, and avoids approximation schemes.
Bolt.jl\footnote{\url{https://github.com/xzackli/Bolt.jl}} by \citet{liBoltjl2023} in Julia is also approximation-free, supports forward-mode automatic differentiation, and uses similar differential equations solvers as SymBoltz.
DISCO-EB\footnote{\url{https://github.com/ohahn/DISCO-EB}} by \citet{hahnDISCODJDifferentiableEinsteinBoltzmann2024} in the JAX framework in Python is also differentiable and relaxes approximation schemes to avoid overhead from switching equations on GPUs.
SymBoltz.jl\footnote{\url{https://github.com/hersle/SymBoltz.jl}} is another Boltzmann code that brings together several of these developments.

The core task of an Einstein–Boltzmann solver is to solve the Einstein–Boltzmann equations for some cosmological model.
They are partial differential equations that linearize to ordinary differential equations (ODEs) ${\mathrm{d}\boldsymbol{u}}/{\mathrm{d} \tau} = \boldsymbol{f}(\boldsymbol{u},\boldsymbol{p},\tau)$ for some initial conditions $\boldsymbol{u}(\tau_i)$ and parameters $\boldsymbol{p}$, including several perturbation wavenumbers.
Any output from Einstein–Boltzmann codes is derived from the solution of these ODEs, such as the matter and CMB power spectra.

However, several properties of the equations complicate this task.
First, the equations separate into computational stages that benefit substantially from being solved sequentially for performance and stability, such as the background, perturbations, and line-of-sight integration stages.
It is common to solve each stage with interpolated input from the previous stage, and joining these can be cumbersome and fragment code.
Second, the set of equations is very long and convoluted.
Ideally, parts of the equations should be easily replaceable to accommodate different submodels for gravity and particle species.
It is hard to organize a code with a suitable modular structure that scales well in model space.
Third, wildly different timescales coexist in the equations, making them extremely stiff and intractable to solve with standard explicit ODE solvers.
This stiffness must be massaged away in the equations by approximations or dealt with numerically by implicit ODE solvers.
Fourth, the number of differential equations is very large, particularly for accurate descriptions of relativistic species.
Typical models with accurate treatment of photons and neutrinos need $O(100)$ equations.
Fifth, the perturbations must be solved for many different wavenumbers $k$, and trade-offs between performance and precision must be made.

To overcome these challenges, most Boltzmann solvers are written in low-level high-performance languages such as C, C++, and Fortran.
They are tightly adapted to the pipeline-like computational structure of the problem (e.g., input $\rightarrow$ background $\rightarrow$ thermodynamics $\rightarrow$ perturbations $\rightarrow \ldots \rightarrow$ output in \citealt{lesgourguesCosmicLinearAnisotropy2011}).
This makes sense for programmers, but does not necessarily provide the simplest interface for modelers.

Traditional codes such as CAMB and CLASS have nonexistent or thin abstraction layers.
To modify them, users must work directly in the low-level numerical code and have a good understanding of its internal structure.
For example, to implement a new species, users must often modify the code in many places: input handling for new parameters, new background equations, new thermodynamics equations, new perturbation equations, joining each of these stages, output handling, and so on.
This leads to fragmented code where changes related to one species are scattered throughout the code.

This structure scales poorly in model space.
As more species and gravitational theories are added, each module is intertwined with code from other physical components.
Even if unused components are deactivated by \enquote{if} statements at runtime, their mere presence in the source code increases its complexity, reduces its readability, and makes it harder to add more models.

We can alleviate this problem to some extent by instead forking the code for modified models, so the main code base is not polluted.
But this only shifts the problem.
Forking duplicates the entire code base, even though only small parts are modified.
Forks are often abandoned and do not receive upstream improvements.
They are also incompatible with each other unless we merge them into one code, which reintroduces the first problem.

The two-language problem amplifies this issue, as data analysis often happens in slower high-level languages such as Python.
This shapes Boltzmann solvers to rigid pipelines that must compute everything at once and avoid interception at all costs, in order to maximize performance in the low-level language before passing output back to the high-level language.
Some features really just post-process the ODE solutions, but are appended to the pipeline even if they are peripheral to the core task of an Einstein–Boltzmann solver.
For example, features such as nonlinear boosting and CMB lensing could be outsourced to smaller and interoperating modular packages, but they often become part of \enquote{all-in-one} Boltzmann codes instead.

These patterns lead to big monolithic Boltzmann solvers
that become increasingly complex as they incorporate more models and features beyond their original core scope.
This complexity has even driven development of specialized AI assistants for CLASS \citep{casasCLAPPCLASSLLM2025}.
While existing Boltzmann solvers are impressively well-engineered and such tools are only helpful, we think these are symptoms of unnecessary complexity.

The Einstein–Boltzmann equations are notoriously stiff \citep{nadkarni-ghoshEinsteinBoltzmannEquationsRevisited2017}.
This property of differential equations means their numerical solution is unstable with standard explicit integrators and requires tiny step sizes, making them hard to solve.
Stiffness can arise when multiple and very different time scales appear in the same problem.
This is very common in cosmology, where particles interact very rapidly in a universe that expands very slowly, particularly in the tightly coupled baryon-photon fluid.
Stiff equations are practically impossible to integrate with explicit solvers and require special treatment.
For a long time, Boltzmann solvers have massaged away stiffness with several approximation schemes%
\footnote{Here, \enquote{approximations} refers to schemes that switch between different equations at different times. It excludes techniques such as multipole truncation and line-of-sight integration, which SymBoltz also uses.}
in the equations \citep[e.g.,][]{doranSpeedingCosmologicalBoltzmann2005,blasCosmicLinearAnisotropy2011a}:
\begin{itemize}
\item tight-coupling approximation (TCA);
\item ultrarelativistic fluid approximation (UFA);
\item radiation streaming approximation (RSA);
\item noncold dark matter fluid approximation (NCDMFA);
\item Saha approximation.
\end{itemize}

They involve switching from one set of equations to another when some control variable measuring the applicability of the approximation crosses a threshold.
This can change the variables in the ODE and require reinitializing the integration.
Approximations enable explicit solvers and can improve both the speed and stability of the solution.
However, they put much more load on modelers to derive and validate several versions of the equations in different regimes.
This process must generally be repeated with modifications to the model, as they can invalidate the approximations or reintroduce stiffness.
They also complicate the numerics, as time series from each ODE solution must be stitched together, each separate ODE system can use different tolerances, ODE integrators must be reinitialized, and so on.

Another way to integrate stiff equations is to use appropriate implicit solvers.
PyCosmo first solved the full stiff Einstein–Boltzmann equations with an implicit integrator, followed by Bolt and DISCO-EB.
CLASS also has an implicit solver, but does not permit disabling the tight-coupling approximation, for example.
Historically, implicit solvers may have been underutilized because they are harder to implement than explicit solvers and have a reputation for being slower,
and due to iterative evolution from initial Boltzmann codes that were centered around approximations.
However, new life has recently been breathed into the field of implicit methods, with the development of new solvers combined with techniques such as automatic and symbolic differentiation that make them more feasible and powerful \citep[e.g.,][]{steinebachConstructionRosenbrockWanner2023,ekanathanFullyAdaptiveRadau2025}.

Derivatives are important in scientific computing and in cosmological applications.
For example, algorithms that optimize likelihoods and MCMC samplers for Bayesian parameter inference can take advantage of derivatives of the likelihood with respect to each parameter to intelligently step in a direction where the likelihood increases.
In machine learning, the same applies when training neural network emulators for cosmological observables by minimizing a scalar loss function of parameters.
Some cosmologies are parameterized as boundary-value problems with the shooting method and use nonlinear root solvers such as Newton's method, which needs Jacobians.
Implicit ODE solvers also use Jacobians to solve for values at the next time step.
Fisher forecasting uses the Hessian (double derivative) of the likelihood with respect to parameters to predict how strong parameter constraints that can be placed by data with some uncertainty.
Boltzmann solvers save time by interpolating spectra computed on coarse grids of $k$ and $l$ to finer grids, which can be made more precise with derivatives with respect to $k$ and $l$.
These are all examples where (applications of) Boltzmann solvers need derivatives. There are at least four ways to compute derivatives.

Manual differentiation is human application of differentiation rules,
but it is limited to simple expressions and by human error.
Symbolic differentiation automates this process with computer algebra systems,
but it is inherently symbolic and cannot differentiate arbitrary programs with control flow, such as conditional statements and loops that depend on numerical values.

Finite differentiation approximates the derivative $f^\prime(x) \approx (f(x+\epsilon/2) - f(x-\epsilon/2)) \,/\, \epsilon$ with a small $\epsilon > 0$ (here using central differences) by simply evaluating the program several times.
This can differentiate arbitrary programs, but it is approximate and introduces the step size $\epsilon$ as a hyperparameter that must be tuned for accuracy and stability. It is a brute-force approach that requires $O(n)$ evaluations ($2n$ using central differences) to compute the gradient of a function $f$ of $n$ variables.
Automatic differentiation \citep[e.g.,][]{griewankEvaluatingDerivatives2008} can be understood by viewing any computer program as a (big) composite function%
\begin{equation}
    \boldsymbol{f} = \boldsymbol{f}_N \circ \boldsymbol{f}_{N-1} \circ \cdots \circ \boldsymbol{f}_2 \circ \boldsymbol{f}_1 = \boldsymbol{f}_N(\boldsymbol{f}_{N-1}( \cdots \boldsymbol{f}_2(\boldsymbol{f}_1)))
\label{eq:program}
\end{equation}
of elementary operations $\boldsymbol{f}_i: \mathbb{R}^{m_i} \rightarrow \mathbb{R}^{n_i}$ (think of $\boldsymbol{f}_i$ as the $i$-th line of code).
Then it numerically evaluates the chain rule%
\begin{equation}
    \boldsymbol{J} = \boldsymbol{J}_N \cdot \boldsymbol{J}_{N-1} \cdot \cdots \cdot \boldsymbol{J}_2 \cdot \boldsymbol{J}_1
\label{eq:program_derivative}
\end{equation}
through the Jacobian $(\boldsymbol{J}_n)_{ij} = \partial f_{n,i} / \partial f_{n-1,j}$ of every operation to accumulate the derivative of the entire program.
This is numerically exact and free of precision parameters, while usually requiring fewer operations than finite differences.
But it is perhaps less intuitive, harder to implement, and needs the source code of the program \eqref{eq:program} to interpret it in the nonstandard way \eqref{eq:program_derivative}.

Notably, while the function \eqref{eq:program} must be evaluated inside-to-outside (right-to-left),
the chain rule \eqref{eq:program_derivative} is an associative matrix product that can be evaluated in any order.
This generally changes the number of operations and is a more open-ended computational problem.
Forward-mode automatic differentiation seeds $\boldsymbol{J}_1 = \boldsymbol{1}$ (the derivative of the input with respect to itself) and multiplies $\boldsymbol{J}_N (\boldsymbol{J}_{N-1} (\cdots (\boldsymbol{J}_2 \boldsymbol{J}_1)))$ by \enquote{pushing} every column of $\boldsymbol{J}_1$ forward through the product in the same evaluation order as $\boldsymbol{f}$.
Reverse-mode first computes $\boldsymbol{f}$ in a forward pass, then seeds $\boldsymbol{J}_N = \boldsymbol{1}$ (the derivative of the output with respect to itself) and multiplies $(((\boldsymbol{J}_N \boldsymbol{J}_{N-1}) \cdots) \boldsymbol{J}_2) \boldsymbol{J}_1$ by \enquote{pulling} every row of $\boldsymbol{J}_N$ backwards through the product.
This usually makes forward-mode faster when $\boldsymbol{f}: \mathbb{R}^m \rightarrow \mathbb{R}^n$ has more outputs ($n \gg m$), and reverse-mode better when there are more inputs ($m \gg n$).

In practice, both modes are implemented using techniques called operator overloading or source code transformation.
The former specializes every function on a particular \enquote{dual number} type that propagates both the value and derivative of the function \citep[e.g.,][]{revelsForwardModeAutomaticDifferentiation2016}.
The latter analyzes the source code for $\boldsymbol{f}$ and transforms it into another code that computes $\boldsymbol{J}$.
In any case, automatic differentiation does not work on compiled binaries, but requires access to the source code to compute gradients with a different path through the instructions of the program.

\begin{figure*}
    \centering
    \begin{tikzpicture}[
        remember picture,
        node distance=6.5cm,
        every node/.style = {align=center},
        wrapper/.style = {rectangle, rounded corners, fill=black!10, font=\normalsize, minimum width=4.7cm, minimum height=6.0cm},
        comp/.style = {draw, circle, fill=gray, minimum size = 1.3cm, inner sep = 0pt, font=\tiny},
        interaction/.style = {{Latex[width=1mm,length=1mm]}-{Latex[width=1mm,length=1mm]}},
        stage/.style = {draw, minimum height=1cm, minimum width = 3cm, fill=lightgray!75},
        process/.style = {-{Latex[width=2mm,length=2mm]}, very thick},
        bigprocess/.style = {-{Latex[width=4mm,length=4mm]}, very thick},
        title/.style = {anchor=north, text width=4cm, align=center, yshift=-0.15cm, font=\bfseries},
    ]
    \node (model) [wrapper] {
        \begin{tikzpicture}[remember picture, node distance=1cm]
            \node [comp, fill=black!40] (grav) {Gravity} [interaction, grow cyclic, level 1/.append style = {level distance = 1.7cm, sibling angle = 60}]
            child { node[comp, fill=blue!50] (bar) {Baryons}}
            child { node[comp, fill=magenta!50] (pho) {Photons}}
            child { node[comp, fill=red!50] (neu) {Massless\\neutrinos}}
            child { node[comp, fill=orange!50] (mneu) {Massive\\neutrinos}}
            child { node[comp, fill=green!50] (cdm) {Cold dark\\matter}}
            child { node[comp, fill=cyan!50] (cc) {Cosmo-\\logical\\constant}};
            \draw[interaction] (bar) -- (pho);
        \end{tikzpicture}
    };
    \node[title] at (model.north) {Symbolic model};
    \node (problem) [right of=model, wrapper] {
        \begin{tikzpicture}[font=\small, node distance=1.6cm]
            \node (bg) [stage] {Background: $\boldsymbol{f}$, $\boldsymbol{J}$};
            \node (pt) [stage, below of=bg] {Perturbations: $\boldsymbol{f}$, $\boldsymbol{J}$};
            \node (more) [stage, below of=pt] {\ldots};
            \draw [process] (bg) -- (pt);
            \draw [process] (pt) -- (more);
        \end{tikzpicture}
    };
    \node[title] at (problem.north) {Numerical problem};
    \node (solution) [right of=problem, wrapper] {
        \begin{tikzpicture}[font=\small, node distance=1.6cm]
            \node (bg) [stage] {Background: $S(\tau)$};
            \node (pt) [stage, below of=bg] {Perturbations: $S(\tau, k)$};
            \node (more) [stage, below of=pt] {\ldots};
            \draw [process] (bg) -- (pt);
            \draw [process] (pt) -- (more);
        \end{tikzpicture}
    };
    \node[title] at (solution.north) {Solution object};
    \draw [bigprocess] (model) -- node [above=1mm] {Compile} (problem);
    \draw [bigprocess] (problem) -- node [above=1mm] {Solve} (solution);
    \end{tikzpicture}
    \caption{SymBoltz represents cosmological models with symbolic equations grouped in physical components for the metric, gravity, and particle species. This is compiled to a numerical problem that splits equations into background and perturbation stages and generates fast code for ODE functions $\boldsymbol{f}$ and Jacobians $\boldsymbol{J}$. The problem is then solved, and the result is stored in a solution object that gives access to any model variable $S$.}
    \label{fig:components}
\end{figure*}

\section{Code architecture and main features}
\label{sec:features}

SymBoltz is designed around three main features.
In brief, it has a symbolic-numeric abstraction interface where users enter high-level symbolic equations that are automatically compiled to fast low-level numerical functions.
It cures stiffness with implicit ODE solvers instead of approximation schemes to keep models simple, elegant, and extensible.
It is differentiable, so it is possible to get accurate derivatives of any output with respect to any input.
This provides rapid model prototyping, helpful abstractions, and automates tasks that must be done manually when modifying other codes.
SymBoltz encourages interactive use and pursues a modular design that lets users integrate what they need from the package into their own applications.
The end goal is to prioritize the modeler, who should be able to just write down their equations, while SymBoltz automates modeling chores.

The code is written in the Julia programming language \citep{bezansonJuliaFreshApproach2017a}, which has a rich ecosystem of scientific packages and aims to resolve the two-language problem.
The symbolic-numeric interface is built on ModelingToolkit.jl\footnote{\url{https://github.com/SciML/ModelingToolkit.jl}} by \citet{maModelingToolkitComposableGraph2022} and Symbolics.jl\footnote{\url{https://github.com/JuliaSymbolics/Symbolics.jl}} by \citet{gowdaHighperformanceSymbolicnumericsMultiple2022}.
The compiled functions are solved by implicit ODE integrators in OrdinaryDiffEq.jl\footnote{\url{https://github.com/SciML/OrdinaryDiffEq.jl/}} by \citet{rackauckasDifferentialEquationsjlPerformantFeatureRich2017} and linear algebra methods in LinearSolve.jl\footnote{\url{https://github.com/SciML/LinearSolve.jl}}.
Automatic differentiation works through ForwardDiff.jl\footnote{\url{https://github.com/JuliaDiff/ForwardDiff.jl}} by \citet{revelsForwardModeAutomaticDifferentiation2016}.

The next subsections describe SymBoltz' three main features in depth.
Other implementation details are given in Appendix \ref{sec:implementation}.
This paper describes SymBoltz version 1.0.0.
We refer to the package documentation for definitive up-to-date information.

\subsection{Symbolic-numeric interface}
\label{sec:symbolicnumeric}

The core of SymBoltz is built around a symbolic-numeric interface illustrated in Fig. \ref{fig:components}.
A model defines what equations should be solved; a problem decides how they are solved; and a solution holds what was solved.
Variables and equations are specified in a high-level user-friendly symbolic modeling language, then compiled to low-level numerical code that is integrated by ODE solvers.
With knowledge of the symbolic equations, SymBoltz analyzes their structure programmatically to ease model specification, automates mechanical boilerplate tasks, and generates fast and stable code that avoids approximations.
In our opinion, there are three properties that motivate such a symbolic abstraction layer around the Einstein–Boltzmann equations.

First, the equations are most easily written as one large ODE, but optimally solved as several ODEs in stages such as the background, perturbations, and line-of-sight integration.
Mathematically, the Jacobian matrix of the single ODE encodes which variables are independent and belong to which stages.
It is easiest for modelers to specify equations in one common system, where the perturbations can refer to variables in the background, and then separate the computational stages automatically.

Second, solving each stage is hard due to stiffness, performance requirements, and dependence on previous stages.
To overcome this, it helps to automatically generate ODE code that is optimal in speed, accuracy, and stability, generate the ODE Jacobian needed by implicit solvers, and interpolate variables from previous stages.
Modelers should not have to do this manually.

Third, the equations are often modified, as we do not know the true cosmology.
This can be made easier by organizing equations in a way that reflects the physical structure of the model, so it is easy to replace one related subset of equations with another without duplicating the entire model.
This could be slow to do at the numerical level, but can be done with no runtime penalty in an intermediate symbolic representation of the equations.

Inspired by this, SymBoltz inverts the traditional layout of Boltzmann codes.
Most codes are built as pipelines following the background, perturbations, and other stages,
but SymBoltz is primarily structured around the physical components that make up the equations.
Related variables and equations are grouped in distinct components for the metric, gravity, photons, baryons, dark matter, dark energy, neutrinos, and other species (see Fig. \ref{fig:components}).
This isolates everything related to one submodel in one place.
Any set of such submodels is joined into a full cosmological model, such as \LCDM{}.
Interactions (e.g., Compton scattering or sourcing of gravity) are equations that connect components.
Adding new submodels is easy, and SymBoltz is largely devoted to simply building a well-organized library of them.

SymBoltz then takes a full cosmological model, separates it into stages, and generates numerical code to solve each stage.
This preprocessing does not slow the code at runtime.
To the contrary, the symbolic equations give extra information such as the Jacobian that enhances speed and stability.
Modifications automatically enjoy these benefits without extra user input.

The modular component-based structure scales well in model space.
For example, modified gravity or dark energy models can be written as self-contained submodels and combined with other components to create many different extended cosmological models.
The generic symbolic framework also handles reduced models with noninteracting radiation, matter and dark energy species.
When exploring modified gravity theories, it can be helpful to test such toy models to understand how gravity responds to pure fluids without coupling to baryons and photons.

In contrast, the monolithic layout of traditional codes scatters changes for one submodel across different modules for input, the background, perturbations, output, and so on.
As more submodels are added, the code is increasingly intertwined and fragmented.
They are also not flexible enough to test simpler toy models, as baryons and photons are inherently hardcoded.
This design fits the computational procedure better than the physical structure.
It can lead to an overwhelming code that is hard to read and modify.
This complexity scales poorly in model space.

As an alternative to component-based modeling,
SymBoltz also includes a version of the default \LCDM{} model with all equations packed into one big \enquote{unstructured} system.
This makes it trivial to change anything in the equations, but sacrifices modularity.
The symbolic interface makes it very compact:
SymBoltz defines a full \LCDM{} model in only 277 lines of code,%
\footnote{\url{https://hersle.github.io/SymBoltz.jl/stable/LCDM/}}
while the equivalent code to browse in CLASS is spread over 10 files with 27721 lines.%
\footnote{\texttt{input.\{h,c\}}, \texttt{background.\{h,c\}}, \texttt{thermodynamics.\{h,c\}}, \texttt{perturbations.\{h,c\}}, and \texttt{wrap\_recfast.\{h,c\}} counted by \texttt{wc}.}

PyCosmo's symbolic interface also generates code for ODEs and sparse Jacobians,
but does not focus on component-based modeling or automating tasks such as stage separation.
In this way, SymBoltz' interface simplifies the solution and modification of the Einstein–Boltzmann equations.
It aims to maximize speed, stability, and convenience with minimal user input.
The next subsections explain how the symbolic features work in detail,
and Sect. \ref{sec:examples} shows them in action in a concrete example.

\subsubsection{Automatic numerical code generation}
\label{sec:codegen}

SymBoltz automatically compiles symbolic equations to numerical code for ODEs
\begin{equation}
    \frac{\mathrm{d}\boldsymbol{u}}{\mathrm{d} \tau} = \boldsymbol{f}(\boldsymbol{u},\boldsymbol{p},\tau).
    \label{eq:ode}
\end{equation}
The generated code is fast and prevents users unfamiliar with Julia or SymBoltz from writing slow code.
If necessary, we can escape the standard code generation and call arbitrary numerical functions; for example, to solve a nonlinear equation for the minimum of a potential or interpolate tabulated data.
The code generation handles tasks such as allocating indices for each ODE state $u_i$, and does optimizations such as common subexpression elimination (see Sect. \ref{sec:modifying}).
This is helpful as Boltzmann solvers tend to have large $\boldsymbol{f}$, and ODE solvers evaluate $\boldsymbol{f}$ many times.

\subsubsection{Automatic handling of observed variables}
\label{sec:observed}

A general ODE \eqref{eq:ode} has two types of variables: \enquote{unknowns} $\boldsymbol{u}(\tau)$ are integrated with respect to time, while \enquote{observed} variables are any functions of the unknowns.
The Einstein–Boltzmann equations are usually written with many observed variables.

For example, consider the metric and gravity equations in Appendices \ref{sec:metric} and \ref{sec:gravity} sourced by some known $\rho(\tau)$, $\delta\rho(\tau,k)$, and $\Pi(\tau,k)$.
Here, $a(\tau)$ and $\Phi(\tau,k)$ are the only unknown (differential) variables that the ODE is integrated for,
whereas $z(\tau)$, $\scrH(\tau)$, $H(\tau)$, and $\Psi(\tau,k)$ are all observed (algebraically) from the unknowns.
Furthermore, it is not always straightforward to express observed derivatives such as $\scrH^\prime$ or $\Psi^\prime$, but SymBoltz automatically expands them using the definitions of $\scrH$ or $\Psi$.
Of course, we can eliminate all observeds by explicitly inserting them into the equations for the unknowns.
However, this reduces readability, as observed variables are helpful intermediate definitions that break up the equations, and we may want to extract them from the solution as well.
Furthermore, modified models can change the sets of unknown and observed variables (e.g., modified gravity can change the constraint equation for $\Psi$ into a differential equation).
It is easier to build models when variables are not hardcoded as either unknown or observed.

SymBoltz reads a full system of equations as input, such as that defined by the entirety of Appendix \ref{sec:implementation}, and automatically separates unknown and observed variables.
After solving the ODE for its unknowns, SymBoltz can automatically recompute any observed variable from its expansion in unknowns.

The bottom line is that users can easily use any variable anywhere by just referring to it, whether it is unknown or observed.
In other solvers it can be necessary to look up observed expressions and recompute them manually.
This is tedious, error-prone, and can tempt users to unnecessarily make simplifying assumptions (e.g., approximating $\Psi \approx \Phi$ without anisotropic stress).

\subsubsection{Automatic stage separation and splining of unknowns}
\label{sec:splining}

In principle, the entire Einstein–Boltzmann system (i.e., background and perturbations) can be integrated all at once.
However, it can be broken down into sequential computational stages that each depend only on those before it.
To alleviate stiffness in each stage, avoid recomputing the background for every perturbation mode, and to improve performance by integrating smaller ODEs, all Boltzmann codes solve the system stage-by-stage and spline variables from one stage as input to the next.

To illustrate this case, again consider the general relativistic equations in Appendix \ref{sec:gravity} sourced by some known $\rho(\tau)$, $\delta\rho(\tau,k)$, and $\Pi(\tau,k)$.
Clearly, $\Phi(\tau,k)$ and $\Psi(\tau,k)$ depend on $a(\tau)$, but $a(\tau)$ does not depend on $\Phi(\tau,k)$ or $\Psi(\tau,k)$, reflecting the perturbative nature of the problem.
We can first solve for only $a(\tau)$ in the \enquote{background,} then spline and look up $a(\tau)$ to solve for $\Phi(\tau,k)$ and $\Psi(\tau,k)$ in the \enquote{perturbations,} instead of solving for all three together and repeatedly integrate $a(\tau)$ for every $k$.

SymBoltz uses the same stage separation strategy as other Boltzmann codes,
but automates it with knowledge of the symbolic equations.
First, all equations are split into background and perturbation stages.
A Cubic Hermite spline is then constructed for all background unknowns (e.g., $a(\tau)$, $X_\Hy^+(\tau)$, $X_\He^+(\tau)$, $T_b(\tau)$, and $\kappa(\tau)$) and replaces the background unknowns in the perturbations.
One vector-valued spline is used to avoid repeatedly looking up $\tau$ for every unknown.
Hermite splines are optimal for interpolating ODEs, as they take both $\boldsymbol{u}(\tau)$ and $\boldsymbol{u}^\prime(\tau)$ into account for better accuracy, and $\boldsymbol{u}^\prime(\tau)$ is known analytically from $\boldsymbol{u}(\tau)$ through the ODE \eqref{eq:ode}.
In contrast, observed variables (e.g., $\scrH(\tau)$) are computed from the (splined) unknowns, as their derivatives are not known directly and splining them is less accurate.
This makes any background variable available in the perturbations \enquote{for free.}
All stages of the Einstein–Boltzmann equations are solved as if they are one common system of equations,
while stage separation is done automatically under the hood.

The separation into background and perturbation stages is guaranteed.
It just reflects the perturbative structure of the linear Einstein–Boltzmann equations, where each order depends only on lower orders.
However, they can often be broken further down: most thermodynamics (recombination) models can be separated from the background, and integral solutions (e.g., the optical depth $\kappa(\tau) = \int_{\tau_0}^\tau \kappa^\prime(\bar\tau) \mathrm{d}\bar\tau$ and line-of-sight integration) can be computed after solving the differential equations.
Future work could extend SymBoltz' background-perturbations separation to split all finer stages by symbolically inspecting the dependencies between variables in the equations.

\subsubsection{Automatic solution interpolation}
\label{sec:interpolation}

SymBoltz integrates the background and perturbation ODEs in conformal time $\tau$ and for several wavenumbers $k$.
All the results are stored in a dedicated solution object.
This object can be queried for any variable or symbolic expression
\begin{equation}
    S(\tau) \quad \text{or} \quad S(\tau, k).
\end{equation}
This looks up the background solution if $S$ is a background variable, or the perturbation solutions if it is perturbative.
It interpolates from the times stored by the ODE solver to the requested $\tau$ using the solver's dense interpolation scheme.
If $S$ is perturbative, it also interpolates from the solved perturbation $k$-modes to the requested $k$.
If $S$ is an unknown, it is returned directly from the ODE solution.
If $S$ is observed, it is instead recomputed automatically from its expansion in terms of unknowns.

The result is that the user can access any variable in the model without having to do any of this manually.
The solution interpolation is also incorporated into plotting recipes that easily visualize any variable as a function of $\tau$ and $k$.

\subsubsection{Automatic Jacobian generation and sparsity detection}
\label{sec:jacobian}

Just as SymBoltz generates code for $\boldsymbol{f}$ in the ODE \eqref{eq:ode}, it uses the same symbolic equations to generate its Jacobian $\boldsymbol{J}$ with entries
\begin{equation}
    J_{ij} = \frac{\partial f_i}{\partial u_j}.
\label{eq:jacobian}
\end{equation}
Jacobians are crucial for solving stiff ODEs with implicit solvers.
Manually coding them is tiresome, error-prone, and must be repeated for new models.
Numerical evaluation with finite differences is approximate and slow.
Automatic differentiation can compute $\boldsymbol{J}$ from $\boldsymbol{f}$, but makes it harder to exploit the sparsity of $\boldsymbol{J}$.
Analytical Jacobians are as fast and stable as possible,
and SymBoltz generates them automatically for the solver without distracting the modeler from focusing on the physical equations.

SymBoltz stores the Jacobian in sparse form to boost performance when it has many zeros.
From the analytical $\boldsymbol{J}$, SymBoltz precomputes its exact sparsity pattern (where $J_{ij} = 0$).
This is hard to specify manually, and numerical solvers cannot distinguish false local zeros (for some inputs) from true global zeros (for all inputs).
Without approximations, SymBoltz reuses the same sparsity structure throughout the integration, computes the nonzero $J_{ij}$ analytically, and stores them at a precomputed index in the sparse $\boldsymbol{J}$.
This avoids converting $\boldsymbol{J}$ from dense to sparse form at runtime.
Easy analytical and sparse Jacobians are crucial for SymBoltz to remain fast without approximations.
This is a major advantage with the symbolic approach, as in PyCosmo.

\subsection{Approximation-freeness}
\label{sec:approx-free}

SymBoltz treats stiffness in the Einstein–Boltzmann equations with modern implicit ODE solvers that integrate the full equations at all times.
It is therefore free of approximation schemes, such as the TCA, UFA, RSA, NCDMFA, and Saha approximation.
This is friendlier to the modeler, who now has to provide only one set of equations, instead of deriving, implementing, and validating approximations.
This approach is also well-suited to the high-level equation-oriented symbolic interface.

Implicit solvers take more expensive steps than explicit methods.
At every step, they solve a generally nonlinear system of equations for the next unknowns $\boldsymbol{u}$, often using Newton's method.
It iteratively solves linear systems $\boldsymbol{W} \boldsymbol{u} = \boldsymbol{b}$
by LU-factorizing $\boldsymbol{W} = \boldsymbol{I} - \gamma h \boldsymbol{J}$,
where $\gamma$ is a constant, $h$ is the time step, and $\boldsymbol{J}$ is the ODE Jacobian \eqref{eq:jacobian}.%
\footnote{
For example, the implicit Euler method $\boldsymbol{u}_{n+1} = \boldsymbol{u}_n + h \boldsymbol{f}(\tau_{n+1}, \boldsymbol{u}_{n+1})$ solves
$\boldsymbol{F}(\boldsymbol{u}_{n+1}) = \boldsymbol{u}_{n+1} - h \boldsymbol{f}(\tau_{n+1}, \boldsymbol{u}_{n+1}) - \boldsymbol{u}_n = \boldsymbol{0}$ at every step with Newton's method
using its Jacobian $\boldsymbol{W} = \boldsymbol{\nabla} \boldsymbol{F} = \boldsymbol{I} - h \boldsymbol{J}(\tau_{n+1},\boldsymbol{u}_{n+1})$.
}
LU-factorizing a dense $n \times n$ matrix from an $n$-dimensional ODE costs $O(n^3)$ operations
and bottlenecks larger ODEs (e.g., larger $\lmax$).
In return, implicit methods use more information to take long and stable steps.

To be fast, implicit solvers need several optimizations.
They must compute not only $\boldsymbol{f}$, but also $\boldsymbol{J}$ efficiently.
A common trick is to reuse an LU-factorized $\boldsymbol{W}$ over several steps, even if it really changes.
Newton's method only requires an approximate $\boldsymbol{W}$ to find the root,
and $\boldsymbol{W}$ is recomputed only if convergence is slow due to an outdated $\boldsymbol{J}$ or a changed step size $h$.
This trades fewer LU-factorizations and $\boldsymbol{J}$ evaluations for more linear solve back-substitutions, which only cost $O(n^2)$ operations.
The next trick is to speed up the linear algebra with sparse matrices if $\boldsymbol{W}$ has many zeros, as for the perturbations.
SymBoltz handles this by generating the analytical and sparse Jacobian (see Sect. \ref{sec:jacobian}).

SymBoltz generates code for all implicit solvers in OrdinaryDiffEq, and dense and sparse matrix methods in LinearSolve.
We can test different solvers, which is useful as performance and stability of implicit methods and linear algebra operations are problem-dependent.
Using $O(100)$ perturbation equations with more than $95\%$ sparsity, we find \texttt{RFLUFactorization} in LinearSolve fastest for dense matrices, but \texttt{KLUFactorization} several times faster when exploiting sparsity.
We tested these ODE solvers from OrdinaryDiffEq:

\begin{itemize}
\item
Backward differentiation formula (BDF) methods,
such as \texttt{FBDF} and \texttt{QNDF} \citep{shampineMATLABODESuite1997}, are
multi-step methods that use several past points to solve for the next.
They do not have Runge–Kutta \enquote{stages} and solve just one implicit equation per step,
so they scale well to large ODEs and can reuse Jacobians heavily.
They have variable order of 1-5, but are \enquote{L-stable} only up to second order.
This makes them take short steps in stiff regimes without utilizing their high orders.
We find them to be slowest.
The solvers \texttt{ndf15} in CLASS and \texttt{BDF2} in PyCosmo also belong to this family.
\citet{nadkarni-ghoshEinsteinBoltzmannEquationsRevisited2017} also discuss BDF methods in the context of the Einstein–Boltzmann equations.

\item
Explicit singly diagonal implicit Runge–Kutta (ESDIRK) methods \citep[e.g.,][]{jorgensenFamilyESDIRKIntegration2018} are a class of implicit Runge–Kutta methods designed to preserve most stability properties of fully implicit (FIRK) methods at a cheaper computational cost.
An $s$-stage FIRK method has excellent stability, but a dense Butcher tableau $a_{ij}$, and must solve a system of $n \times s$ equations at every step.
D in ESDIRK means that $a_{ij}$ has only lower triangular and diagonal nonzeros, decoupling the system into $s$ systems of $n$ equations that are faster to solve sequentially.
S means that all diagonal nonzeros $a_{ii} = \gamma$ are equal, so all stages share the same $\boldsymbol{W}$-matrix, and only one must be factorized and can be reused as in BDF methods.
E means that $a_{11} = 0$, so the first stage is explicit and cheap.
Unlike BDF methods, they can remain L-stable without losing order to stiffness.
We find the fourth order \texttt{KenCarp4} method \citep{kennedyAdditiveRungeKutta2003} to perform well with much fewer time steps.
Bolt also defaults to it. DISCO-EB uses \texttt{Kvaerno5}, but we find it to be slower.

\item Rosenbrock methods \citep[e.g.,][]{langRosenbrockWannerMethodsConstruction2020} linearize $\boldsymbol{f} \approx \boldsymbol{J} \boldsymbol{u}$ and effectively replace Newton's method with one linear solve.
Being free from nonlinear convergence issues is a major upside and suits the linear perturbation ODEs.
The downside is that they can need a new $\boldsymbol{J}$ at every step, which is slow to evaluate in many problems.
But generating $\boldsymbol{J}$ analytically eliminates this issue.
The subclass of \enquote{Rosenbrock-W} methods can reuse Jacobians, but this strategy is not yet used by OrdinaryDiffEq.
We find the fifth order L-stable \texttt{Rodas5P} method \citep{steinebachConstructionRosenbrockWanner2023} to be both most efficient and stable on both the background and perturbations, and have not found it used by other Boltzmann codes.
\end{itemize}
Section \ref{sec:performance} compares the performance of these methods.

Traditional codes use approximations to reduce stiffness and increase performance by evolving fewer ODE states when possible.
Earlier works find marginal speedups from the TCA and UFA compared to using implicit solvers,
but the RSA can significantly speed up models with high $\lmax$ \citep{lesgourguesCosmicLinearAnisotropy2011a,moserSymbolicImplementationExtensions2022,hahnDISCODJDifferentiableEinsteinBoltzmann2024}.
However, approximation-based codes must account for the structure of the equations to change, which adds overhead to ODE solvers and restructuring of sparse Jacobians.
Approximation-free solvers can challenge this by optimizing around unchanged structure.
Numerical codes approximate $\boldsymbol{J}$ with finite differences using $O(n)$ evaluations of $\boldsymbol{f}$.
This is wasteful, as $\boldsymbol{f} = \boldsymbol{J} \boldsymbol{u}$ for linear perturbations, so computing the nonzeros of $\boldsymbol{J}$ should take fewer operations than $\boldsymbol{f}$!
Symbolic codes can take advantage of such properties.
Approximations also complicate the code and put more load on modelers to derive, implement, and validate them,
and repeat the process as modifications reintroduce stiffness or invalidate approximations.

SymBoltz is tuned for performance without invoking approximations.
The approximation-free structure is a major simplification and pairs well with a symbolic high-level equation-oriented interface.
In the long run, approximation schemes can of course be explored as a secondary and optional way to maximize speed.

\subsection{Differentiability}
\label{sec:diff}

SymBoltz is compatible with automatic differentiation.
It can compute derivatives of any output quantity (e.g., ODE variables, $P(k)$, or $C_l$) with respect to any input parameters (e.g., $h$ or $\Omega_{c0}$).
For example, it can be used to learn the sensitivity of the output to the input, perform Fisher forecasts, or compute the Jacobian in the shooting method for boundary-value parameterized models.
It can also compute the ODE Jacobian, but SymBoltz does this symbolically to get it analytically with its sparsity pattern.

Currently, SymBoltz and Bolt both work only with forward-mode dual numbers through ForwardDiff.
This is appropriate for applications with more outputs than inputs, such as differentiating $O(100)$ values of $P(k)$ or $C_l$ as functions of $O(10)$ parameters.
Reverse-mode is ideal with fewer outputs and attractive for MCMCs or training emulators where output is compressed to a scalar loss.
Fast reverse-mode gradients could accelerate sampling in large parameter spaces from upcoming surveys and challenge the use of emulators.
Notably, DISCO-EB supports reverse-mode, but has not yet demonstrated it for this purpose.
However, reverse-mode in SymBoltz is left for future work.

\section{Examples}
\label{sec:examples}

The features in Sect. \ref{sec:features} are best illustrated through some examples.
This code is as of SymBoltz version 1.0.0 and available in a notebook in the project repository.

\subsection{Basic usage workflow}

\begin{figure*}
    \centering
    \includegraphics[scale=0.425]{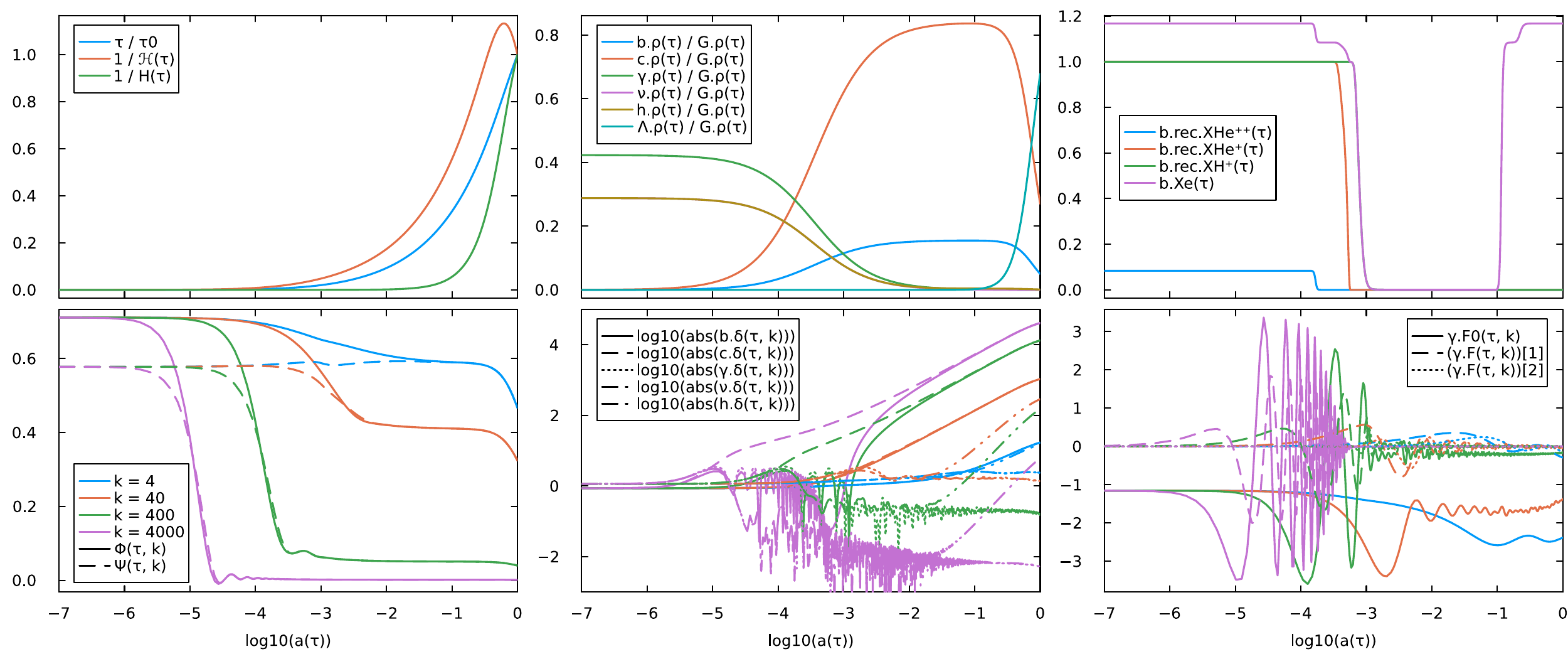}
    \caption{SymBoltz includes plotting recipes that make it easy to visualize any symbolic variable or expression thereof from a solution of the Einstein–Boltzmann equations. This plot was made with one short line of code per subplot. Wavenumbers $k$ are in units of $H_0/c$.}
    \label{fig:output}
\end{figure*}

Most usage follows a model-problem-solution workflow:
\begin{codebox}
\begin{Verbatim}
using SymBoltz
M = ΛCDM(lmax = 16)
p = Dict(
  M.γ.T₀ => 2.7, M.b.Ω₀ => 0.05, M.b.YHe => 0.25,
  M.ν.Neff => 3.0, M.c.Ω₀ => 0.27, M.h.m_eV => 0.02,
  M.I.ln_As1e10 => 3.0, M.I.ns => 0.96, M.g.h => 0.7
)
prob = CosmologyProblem(M, p; jac=true, sparse=true)
ks = [4, 40, 400, 4000] # k / (H₀/c)
sol = solve(prob, ks)
\end{Verbatim}
\end{codebox}
The model-problem-solution split achieves three distinct goals.

First, a symbolic representation \texttt{M} of the \LCDM{} model is created.
This is a standalone object designed to be interactively inspected and modified.
It contains every variable, parameter, and equation of the model structured as one submodel per physical component: the metric $g$, gravitational theory $G$, photons $\gamma$, massless neutrinos $\nu$, massive neutrinos $h$, cold dark matter $c$, baryons $b$, the cosmological constant $\Lambda$, and inflation $I$.
For example, \texttt{equations(M)} shows all its equations; \texttt{equations(M.G)} only the gravitational ones; \texttt{M.g.a} refers to the scale factor variable $a(\tau)$ that \enquote{belongs} to the metric $g$; \texttt{M.b.Ω₀} is the baryon density parameter $\Omega_{b0}$; and \texttt{parameters(M)} shows all parameters that can be set.
Everything displays with \LaTeX-compatibility in notebooks and transparently shows the contents of the model.

Second, the symbolic model is compiled to a numerical problem \texttt{prob} with parameters \texttt{p}.
This expensive step does the processing in Sect. \ref{sec:symbolicnumeric}: it checks equations for consistency; splits them into background and perturbation stages; distinguishes observed and unknown variables; generates fast code for ODEs and Jacobians in numerical or analytical and dense or sparse form; splines background unknowns in the perturbations; and finds initial conditions.
Optional keyword arguments customize the compilation and control precisely how the problem is solved.
This is a separate step because it does final transformations on the model \texttt{M} once the user has finished modifying and committed to it.

Third, \texttt{solve(prob, ks)} solves the background and perturbations for the wavenumbers \texttt{ks}.
Omitting \texttt{ks} solves only the background.
The resulting solution object \texttt{sol} provides convenient access to all model variables.
Internally, it stores values of all ODE unknowns at the time steps taken by the solvers for every requested wavenumber.
However, it can be queried with any time, wavenumber, and symbolic expression, and will automatically compute it from the unknowns and interpolate between stored times and wavenumbers, as explained in Sect. \ref{sec:interpolation}.
For example, \texttt{sol(g.Φ+g.Ψ, 1.0, 2.0)} computes $\Phi+\Psi$ at $\tau = 1 \, H_0^{-1}$ and $k = 2 \, H_0/c$ by expanding $\Psi$ in terms of $\Phi$ and other unknowns.
This interface is also used to plot the evolution of several variables in Fig. \ref{fig:output} with very compact code.

Most Boltzmann solvers save only a fixed set of variables, such as only the unknowns.
Recovering observed variables requires manual effort.
This is cumbersome, open to user error, and adds friction in the modeling process.
SymBoltz is designed to provide easy access to all variables defined by the model.

\subsection{Modifying models}
\label{sec:modifying}

Suppose we want to replace the cosmological constant $\Lambda$ with another dark energy model,
such as dynamical $w_0 w_a$ dark energy \citep{chevallierAcceleratingUniversesScaling2001,linderExploringExpansionHistory2003} with equation of state%
\begin{equation}
    w(\tau) = w_0 + w_a (1-a(\tau)).
\end{equation}
In this case, it is possible to solve the continuity equation
\begin{subequations}
\begin{equation}
\rho^\prime(\tau) = -3\scrH(\tau)\rho(\tau)(1+w(\tau))
\label{eq:w0wadiff}
\end{equation}
analytically with the ansatz $\rho(\tau) \propto a(\tau)^m \exp (n a(\tau))$ to get
\begin{equation}
    \rho(\tau)=\rho(\tau_0)a(\tau)^{-3(1+w_0+w_a)}\exp(-3w_a(1-a(\tau))).
\label{eq:w0waanal}
\end{equation}
\end{subequations}
Perturbations are given by \citet{putterMeasuringSpeedDark2010}.
To implement this species in SymBoltz, we write down all related variables, parameters, equations, and initial conditions in one place:
\begin{codebox}
\begin{Verbatim}
g, τ, k = M.g, M.τ, M.k
a, ℋ, Φ, Ψ = g.a, g.ℋ, g.Φ, g.Ψ
D = Differential(τ)

@parameters w₀ wₐ cₛ² Ω₀ ρ₀=3Ω₀/8π
@variables ρ(τ) P(τ) w(τ) cₐ²(τ) δ(τ,k) θ(τ,k) σ(τ,k)
eqs = [
  w ~ w₀ + wₐ*(1-a)
  ρ ~ ρ₀ * a^(-3(1+w₀+wₐ)) * exp(-3wₐ*(1-a))
  P ~ w * ρ
  cₐ² ~ w - 1/(3ℋ) * D(w)/(1+w)
  D(δ) ~ 3ℋ*(w-cₛ²)*δ - (1+w) * (
         (1+9(ℋ/k)^2*(cₛ²-cₐ²))*θ - 3*D(Φ))
  D(θ) ~ (3cₛ²-1)*ℋ*θ + k^2*cₛ²*δ/(1+w) + k^2*Ψ
  σ ~ 0
]
initialization_eqs = [
  δ ~ -3//2 * (1+w) * Ψ
  θ ~ 1//2 * (k^2*τ) * Ψ
]
X = System(eqs, τ; initialization_eqs, name=:X)
\end{Verbatim}
\end{codebox}

The last line packs everything into a $w_0 w_a$ component \texttt{X}.
Unicode symbols are encouraged to maximize similarity with equations (e.g., \texttt{$\Omega_0$} over \texttt{Omega\_0}),
and SymBoltz automatically renders them in \LaTeX{}-compatible environments (e.g., notebooks).
A full $w_0 w_a \text{CDM}$ model is now built by replacing the cosmological constant species \texttt{$\Lambda$} in $\Lambda \text{CDM}$ by the $w_0 w_a$ species \texttt{X}:
\begin{codebox}
\begin{Verbatim}
M = ΛCDM(Λ=X, lmax=16, name=:w₀wₐCDM)
push!(p, X.w₀ => -0.9, X.wₐ => 0.2, X.cₛ² => 1.0)
prob = CosmologyProblem(M, p; jac=true, sparse=true)
\end{Verbatim}
\end{codebox}

The modification consists of writing down the equations compactly and directly.
This is all the user must do, and under the hood SymBoltz will:
\begin{itemize}
\item create input hooks for setting the parameters $w_0$, $w_a$, $c_s^2$, $\Omega_{0}$;
\item move $(\tau)$-dependent functions to the background;
\item move $(\tau,k)$-dependent functions to the perturbations;
\item expand $\texttt{D(w)} = \mathrm{d}w/\mathrm{d}\tau$ and $\texttt{D($\Phi$)} = \mathrm{d}\Phi/\mathrm{d}\tau$ from $w$ and $\Phi$;
\item compute needed background variables in the perturbations;
\item spline $\rho(\tau)$ in the perturbations (if we use Eq. \eqref{eq:w0wadiff} over \eqref{eq:w0waanal});
\item source gravity with energy-momentum contributions;
\item assign $\Omega_{X0} = 1 - \sum_{s \neq {X}} \Omega_{s0}$ by default (if gravity is GR);
\item eliminate common subexpressions (e.g., $x\hspace{-2pt}=\hspace{-2pt}1\hspace{-1pt}+\hspace{-1pt}w$ and $y\hspace{-2pt}=\hspace{-2pt}1\hspace{-1pt}/\hspace{-1pt}x$);
\item generate ODE functions and state indices for $\delta^\prime$ and $\theta^\prime$;
\item generate analytical and sparse Jacobian entries for $\delta^\prime$ and $\theta^\prime$;
\item interpolate and output any variable from the solution, both for unknowns (i.e., $\delta$ and $\theta$) and observeds (e.g., $w$ and $c_a^2$).
\end{itemize}

More manual work is required in a purely numerical code.
For example, in CLASS, we would have to
read new parameters in \texttt{input.c};
declare new background and perturbation variables in \texttt{background.h} and \texttt{perturbations.h};
solve background equations, save desired output and source gravity in \texttt{background.c};
and recompute or look up background variables, solve perturbation equations, save desired output, and source gravity in \texttt{perturbations.c}.
The changes are scattered across the code, so it grows in complexity as more models are added.
SymBoltz enables a workflow that implements and analyzes a modified model with less effort all in one notebook, for example.
Of course, both codes already include a $w_0 w_a$ species, but this is how they are implemented.
We proceed with the $w_0 w_a \text{CDM}$ model to the next section.

\begin{figure*}[ht!]
    \centering
    \includegraphics[scale=0.425]{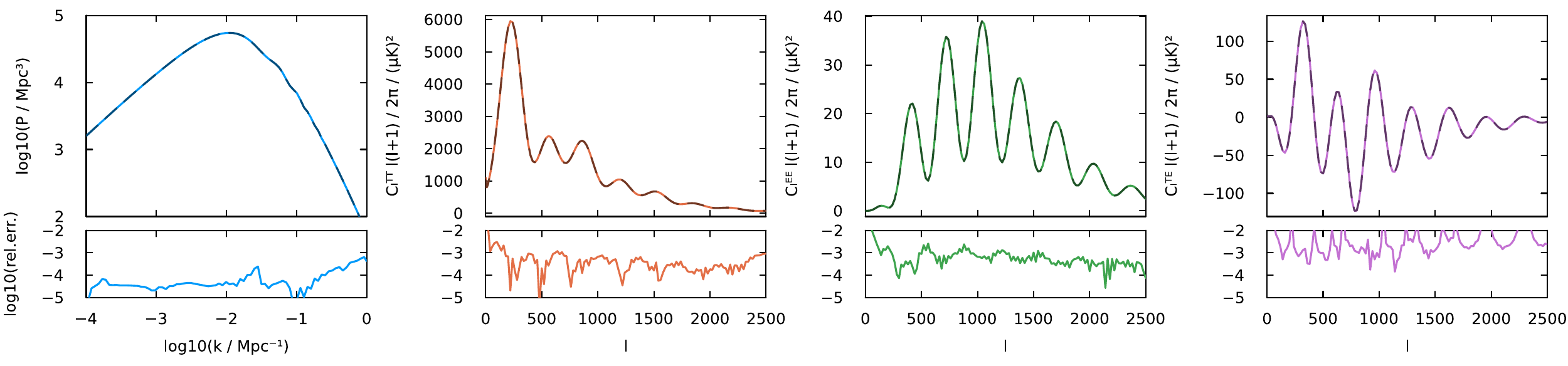}
    \caption{Matter and CMB (TT, EE, and TE) power spectra computed by SymBoltz (colored lines) compared to CLASS (gray dashes) with relative errors $P(k)_\text{SymBoltz}/P(k)_\text{CLASS}-1$ and $C_{l,\text{SymBoltz}}/C_{l,\text{CLASS}}-1$ for the $w_0 w_a \text{CDM}$ model. CLASS uses the precision parameters in Appendix \ref{sec:precision}.}
    \label{fig:spectra}
\end{figure*}

\subsection{Computing power spectra}

\begin{table}[b]
\caption{\label{tab:timings}Time to compute the power spectra in Fig. \ref{fig:spectra}.}
\begin{tabularx}{\linewidth}{X r r}
\toprule
Code (approximations) & $P(k)$ & $C_l$ \\
\midrule
SymBoltz (no approximations) & 0.3 s & 3.1 s \\
CLASS (only tight-coupling approximation) & 5.2 s & 8.7 s \\
CLASS (all approximations) & 0.5 s & 1.6 s \\
\bottomrule
\end{tabularx}
\end{table}

SymBoltz can compute the matter power spectrum $P(k)$, and the angular CMB power spectra $C_l^\text{TT}$, $C_l^\text{EE}$, and $C_l^\text{TE}$:
\begin{codebox}
\begin{Verbatim}
ks = 10 .^ range(-1, 4, length = 100)
Ps = spectrum_matter(prob, ks)
ls = vcat([2,3,5,10], 20:20:2500)
jl = SphericalBesselCache(ls)
Cls = spectrum_cmb([:TT, :EE, :TE], prob, jl)
\end{Verbatim}
\end{codebox}
The \texttt{spectrum\_matter} and \texttt{spectrum\_cmb} functions select a coarse $k$-grid to solve the perturbations for and interpolate to a finer $k$-grid.
This is a common trick to integrate fewer $k$-modes.
Precision parameters that affect the output are given with optional keyword arguments.
Output spectra and their computation time with standard precision settings are shown in Fig. \ref{fig:spectra} and Table \ref{tab:timings}.

It shows that SymBoltz and CLASS agree to around $0.1\%$ or better.
The $P(k)$ agree to $0.1\%$ for all $k$ and $0.01\%$ for linear $k$.
The $C_l$ agree to $0.1\%$ for most $l$, although slightly worse for cosmic variance-dominated $l$ and very large $l$ in $C_l^\text{TE}$.
This is sufficient for most current data \citep[e.g.,][]{bollietHighaccuracyEmulatorsObservables2024}.
It is similar to CAMB versus CLASS with standard precision,
although their agreement is pushed to $0.01\%$ with high precision \citep{lesgourguesCosmicLinearAnisotropy2011c}.
SymBoltz computes $P(k)$ efficiently, but the extra steps needed for $C_l$ are not yet as optimized as they are in CLASS.
Section \ref{sec:performance} gives a more thorough performance comparison for $P(k)$.

\subsection{Differentiable Fisher forecasting}
\label{sec:fisher}

Fisher forecasting is a technique to predict how strong parameter constraints that can be placed by data with given errors.
It needs derivatives that can be computed with automatic differentiation.

Near a peak $\bar{\boldsymbol{p}}$, where derivatives vanish, a log-likelihood $\log L$ of parameters $\boldsymbol{p}$ is approximated by the Taylor series
\begin{equation}
    \log L(\boldsymbol{p}) \approx \log L(\bar{\boldsymbol{p}}) - \sum_{i,j} F_{ij}(\bar{\boldsymbol{p}}) (p_i - \bar{p}_i) (p_j - \bar{p}_j),
\label{eq:likelihood_expansion}
\end{equation}
where $\boldsymbol{F}$ is the Fisher information matrix with elements
\begin{equation}
    F_{ij}(\boldsymbol{p}) = -\frac12 \frac{\partial^2 \log L(\boldsymbol{p})}{\partial p_i \, \partial p_j} .
\label{eq:fisher_general}
\end{equation}
Intuitively, $\boldsymbol{F}$ measures how sharp the peak is or how sensitive $L$ is in different directions in parameter space.
Fisher forecasting is powered by the Cramér–Rao bound $\big\lvert \smash{C_{ij}} \big\rvert \geq \big\lvert \smash{F^{-1}_{ij}} \big\rvert $ for the covariance $C_{ij}$ between parameters $p_i$ and $p_j$.
It is an equality for a Gaussian, for which the likelihood expansion \eqref{eq:likelihood_expansion} is exact.
Thus, inverting $\boldsymbol{F}(\bar{\boldsymbol{p}})$ gives the tightest possible parameter constraints.

To demonstrate differentiable Fisher forecasting with SymBoltz, we make the best possible CMB (TT) measurement of $\bar{C}_l$ over the full sky with errors only due to cosmic variance
\begin{equation}
    \sigma_l = \sqrt{\frac{2}{2l+1}} \bar{C}_l.
\label{eq:cmb_cosmic_variance}
\end{equation}
Assuming a $\chi^2$-log-likelihood
\begin{equation}
    \log L(\boldsymbol{p}) = -\frac12 \sum_l \bigg( \frac{C_l(\boldsymbol{p}) - \bar{C}_l}{\sigma_l} \smash{\bigg)^2} ,
\end{equation}
the Fisher matrix \eqref{eq:fisher_general} becomes
\begin{equation}
    F_{ij} = \sum_l \frac{\partial C_l}{\partial p_i} \frac{1}{\sigma_l^2} \frac{\partial C_l}{\partial p_j} .
\label{eq:fisher_special}
\end{equation}

\begin{figure}
    \centering
    \includegraphics[scale=0.425]{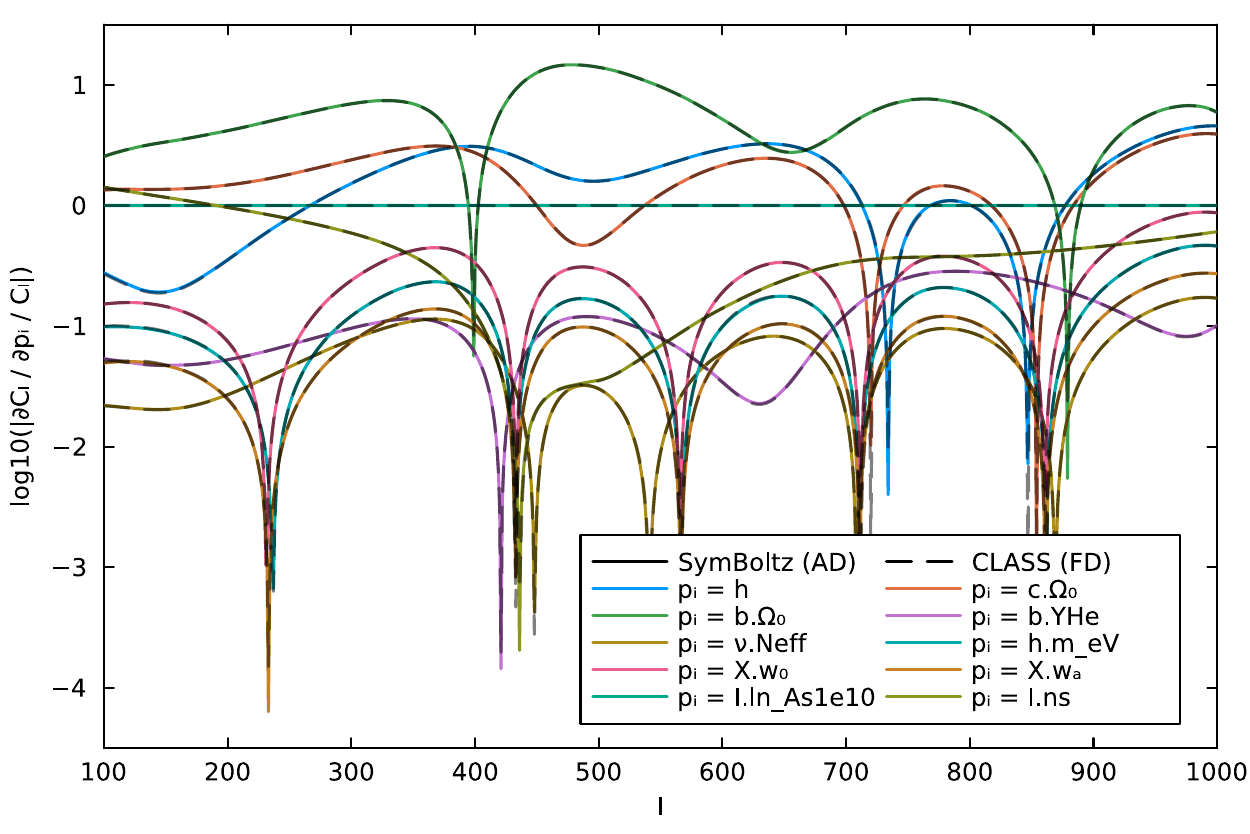}
    \caption{Normalized derivatives $(\partial C_l / \partial p_i) / C_l$ of a CMB TT power spectrum with respect to cosmological parameters $p_i$ from SymBoltz and automatic differentiation (AD; colored lines) versus CLASS and central finite differences (FD; gray dashes). CLASS uses the precision parameters in Appendix \ref{sec:precision} and finite differences with $5 \%$ relative step sizes.}
    \label{fig:derivatives}
\end{figure}

\begin{figure}
    \centering
    \includegraphics[scale=0.425]{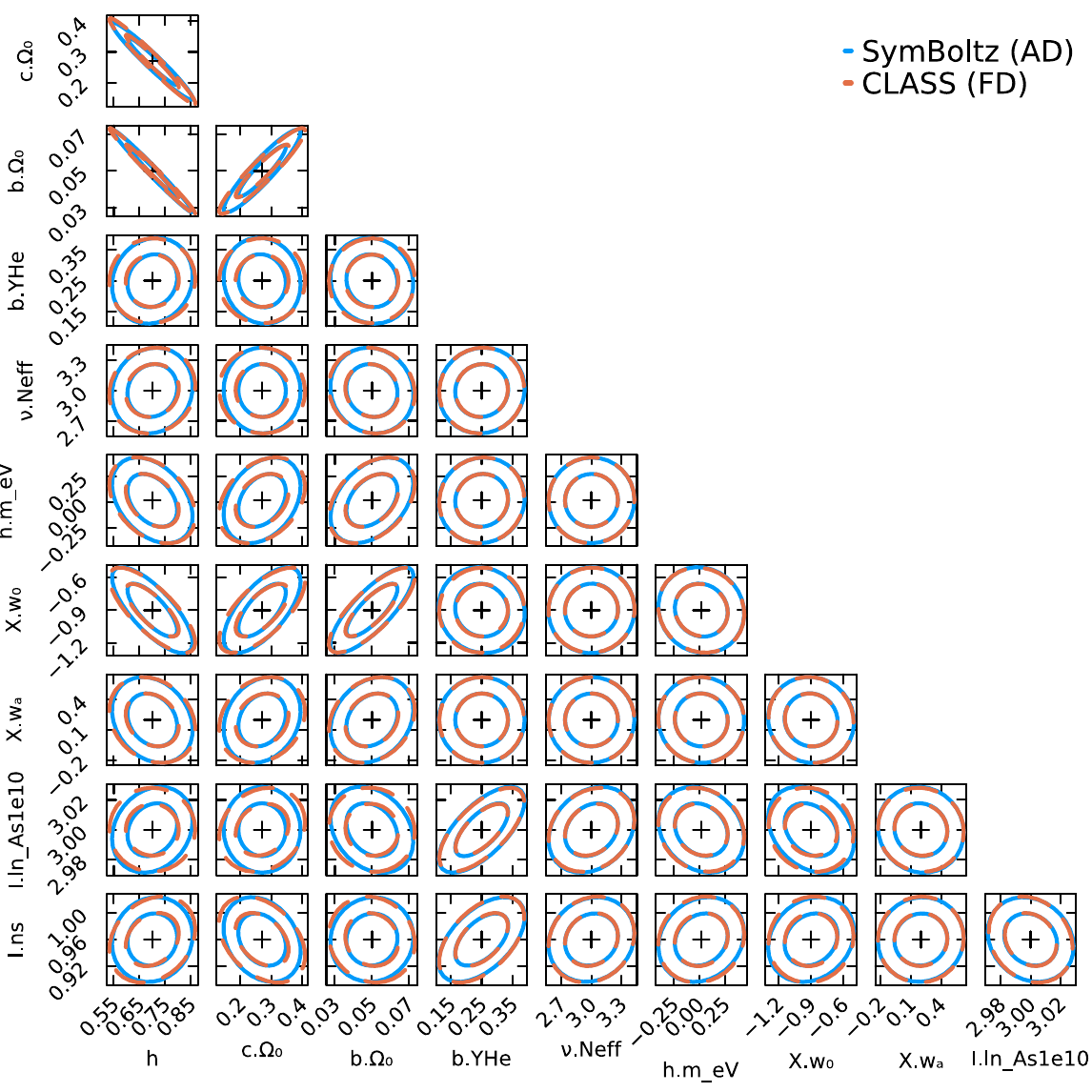}
    \caption{Marginalized 68\% and 95\% 2D confidence ellipses for parameter constraints from a Fisher forecast on a cosmic variance-dominated CMB TT-only survey using the derivatives in Fig. \ref{fig:derivatives}.}
    \label{fig:forecast}
\end{figure}

The derivatives $\partial C_l / \partial p_i$ are usually found with error-prone finite differences and careful step size tuning.
SymBoltz avoids this problem altogether with automatic differentiation:
\begin{codebox}
\begin{Verbatim}
vary = [
  M.g.h, M.c.Ω₀, M.b.Ω₀, M.b.YHe, M.ν.Neff,
  M.h.m_eV, M.X.w₀, M.X.wₐ, M.I.ln_As1e10, M.I.ns,
]
genprob = parameter_updater(prob, vary)
jl, ls = SphericalBesselCache(100:25:1000), 100:1000
Cl(p) = spectrum_cmb(:TT, genprob(p), jl, ls)
p₀ = map(par -> p[par], vary)
dCl_dp = ForwardDiff.jacobian(Cl, p₀)
\end{Verbatim}
\end{codebox}
Here, \texttt{vary} orders the parameters to differentiate with respect to,
and \texttt{genprob} generates a new problem with updated parameters \texttt{p}.
This prepares a vector-to-vector function \texttt{Cl(p)} that is differentiated by \texttt{ForwardDiff.jacobian}
to compute $\partial C_l / \partial p_i$ with dual numbers, as shown in Fig. \ref{fig:derivatives}.
Computing and inverting the Fisher matrix \eqref{eq:fisher_special} forecasts the constraints in Fig. \ref{fig:forecast}.
They agree with finite difference results from CLASS, but these required significant tuning of precision parameters and step sizes.
Differentiable Fisher forecasts have also been demonstrated with an Eisenstein–Hu transfer function fit by \citet{campagneJAXCOSMOEndtoEndDifferentiable2023},
and through the Einstein–Boltzmann solver DISCO-EB by \citet{hahnDISCODJDifferentiableEinsteinBoltzmann2024}.

\subsection{Differentiable MCMC sampling with supernova data}

Finally, SymBoltz can output differentiable results for gradient-based MCMC samplers, such as the No-U-Turn Sampler (NUTS).
It uses gradients to sample parameter space more efficiently than the random walking Metropolis–Hastings algorithm.

We fit 1048 Pantheon observations $(z_i, m_i)$ of Type Ia supernovae by \citet{scolnicCompleteLightcurveSample2018} to predicted apparent magnitudes $m(z) = M + \mu(z)$ at redshift $z$.
Their standard absolute magnitude is $M \approx -19.3$, and $\mu(z) = 5 \lg (d_L(z) / 10\text{ pc})$ is the distance modulus of the background-derived luminosity distance%
\footnote{This expression is valid for any $\Omega_{k0}$ with complex $\sinc(x) = \sin(x)/x$, but can be split into branches for positive, negative, and zero $\Omega_{k0}$.}
\begin{equation}
    d_L(z) = \frac{c}{H_0} \frac{\chi(z)}{a(z)} \sinc \Big( H_0 \sqrt{-\Omega_{k0}} \, \chi(z) \Big) .
\end{equation}

\begin{figure}
    \centering
    \includegraphics[width=1.00\linewidth]{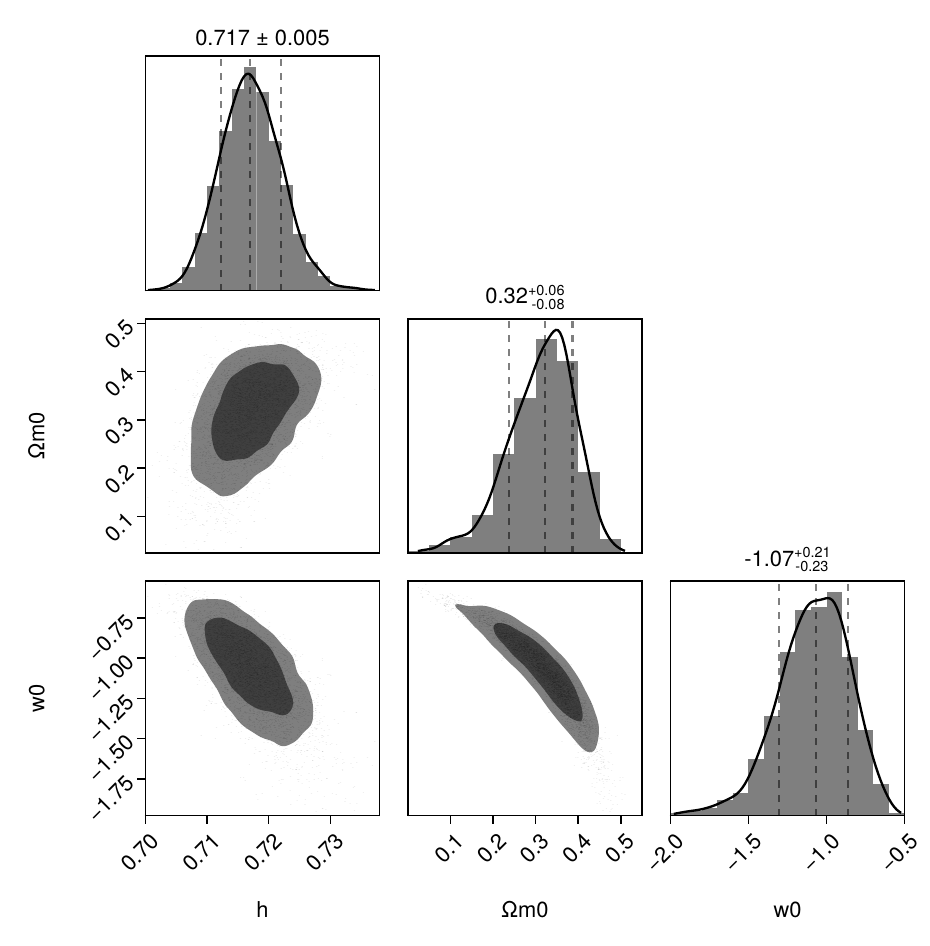}
    \caption{Parameter inference on 1048 Type Ia supernovae from Pantheon data, using 5000 MCMC samples with the gradient-based NUTS sampler in Turing and differentiable predictions from SymBoltz.}
    \label{fig:sn}
\end{figure}

We define the likelihood $L$ by a multivariate normal distribution of the 1048 supernovae with covariance matrix $C$ given by \citet{scolnicCompleteLightcurveSample2018}.
To compute $L$, we use the probabilistic programming framework Turing.jl\footnote{\url{https://github.com/TuringLang/Turing.jl}} by \citet{fjeldeTuringjlGeneralPurposeProbabilistic2025} and specify a probabilistic model equivalent to the log-likelihood%
\begin{equation}
    \log L = -\frac12 \sum_{i,j} C^{-1}_{ij} \big(m(z_i)-m_i\big) \big(m(z_j)-m_j\big) .
\end{equation}
We used the NUTS sampler in Turing.
The code for this example can be found in the paper's notebook.
For the flat $w_0 \text{CDM}$ model ($\Omega_{k0} = 0$, $w_a = 0$, fixed $\Omega_{r0}$), it gives the constraints in Fig. \ref{fig:sn}.

Differentiable computations are not yet fast enough for MCMCs with perturbation-derived spectra.
These preliminary results show that the differentiable pipeline works, and we hope to speed it up with future work.
Differentiable Boltzmann codes have yet to show use with gradient-based MCMC samplers,
but \citet{campagneJAXCOSMOEndtoEndDifferentiable2023} has done more advanced MCMCs with a differentiable background, for example.

\section{Performance comparison to CLASS}
\label{sec:performance}

\begin{figure}
    \centering
    \includegraphics[scale=0.425]{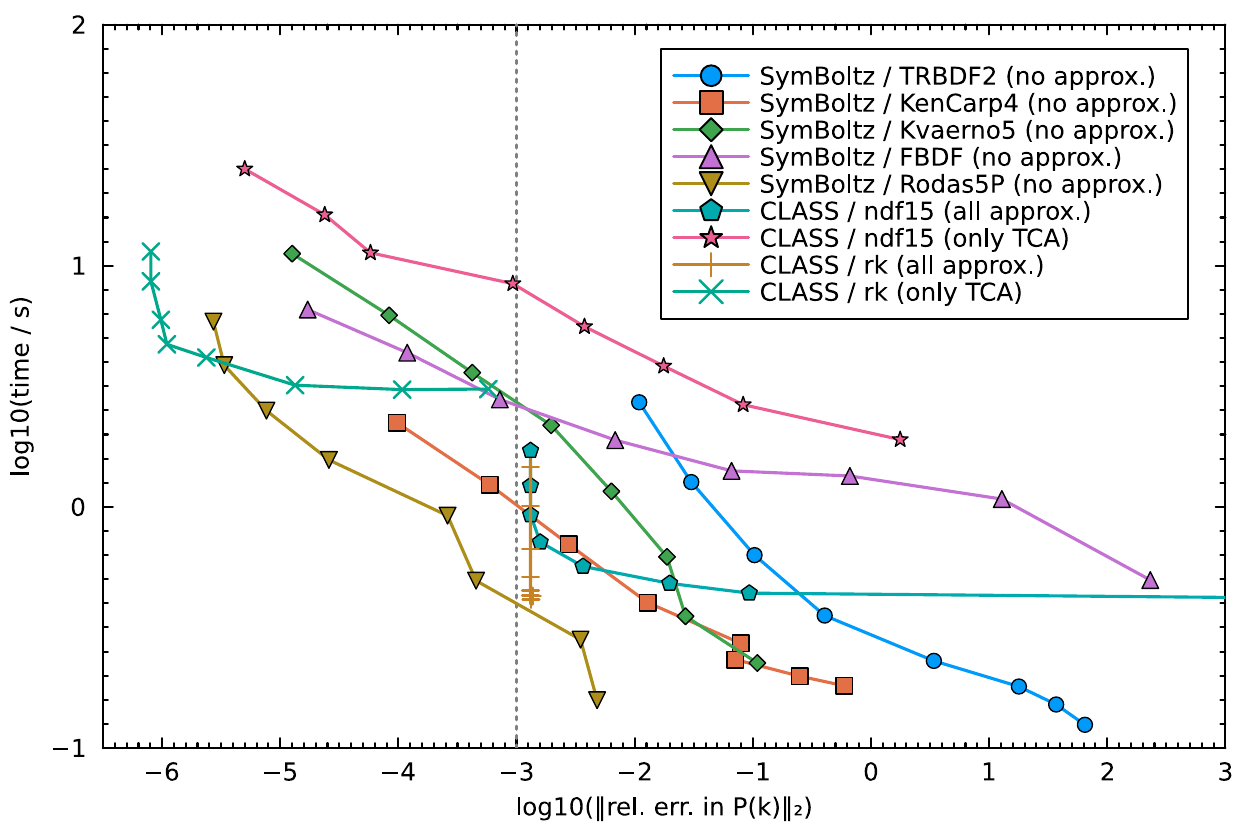}
    \caption{%
        Work versus precision for computing the matter power spectrum $P(k)$ with SymBoltz and CLASS.
        SymBoltz is approximation-free and run with several implicit ODE solvers.
        CLASS is run with approximations on/off, its implicit/explicit evolvers \texttt{ndf15}/\texttt{rk}, and precision settings in Appendix \ref{sec:precision}, but tight coupling is always approximated.
        For each setup, $P(k)$ is computed by integrating perturbation $k$-modes with ODE solver tolerances $10^{-2}$-$10^{-9}$ that controls the adaptive stepping.
        For each tolerance, we record the best of 3 runtimes and the $L_2$-compressed error $\smash{\big(\sum_{i=1}^N \!\big(P(k_i)/\bar{P}(k_i)-1\big)^2/N\big)^{1/2}}$ relative to high-precision spectra $\bar{P}(k)$ from both codes with approximations off, tolerance $10^{-10}$, and \texttt{FBDF}/\texttt{ndf15}.
        Both codes solve a $w_0 w_a \text{CDM}$ model with equal parameters, $N=114$ equal wavenumbers $k_i$, multipole cutoff $\lmax=16$ for all species, default sampling of massive neutrino momenta, and parallelization over $k$.
        Times are from a Linux laptop with an Intel i7-12800H CPU.}
    \label{fig:workprec}
\end{figure}

Figure \ref{fig:workprec} compares performance versus precision of the matter power spectrum $P(k)$ computed with SymBoltz and CLASS.
This computation solves the background, thermodynamics, and perturbations for several $k$ and reads off $P(k)$ at the final time.
It is a relevant test of SymBoltz' approximation-free treatment of the perturbations, as they dominate the computational work.

The comparison shows how approximations and the ODE solver affect the integration of the equations.
Without approximations, SymBoltz with \texttt{Rodas5P} is more than $10\times$ faster than CLASS with its implicit \texttt{ndf15} evolver at comparable precision.
When CLASS uses approximations and its explicit \texttt{rk} evolver, SymBoltz is as fast while remaining approximation-free.

To make this comparison as fair as possible, we configured both codes to integrate identical $k$-modes (decided by CLASS) without $k$-interpolation and use equal multipole cutoffs $\lmax$ for all species.
However, the codes sample massive neutrino momenta $q$ differently.
This greatly impacts the number of perturbation equations (see Appendix \ref{sec:massiveneutrinos}).
SymBoltz' 4 default momenta give 127 perturbation equations in total.
CLASS' default $\texttt{tol\_ncdm\_newtonian} = 10^{-5}$ gives 11 momenta and 245 equations (before approximations).
Increasing this tolerance to $10^{-3}$ gives 5 momenta and 143 equations.
This runs around $245/143 \approx 1.7$ times faster, but introduces errors beyond $1\%$ in $P(k)$.
We therefore leave both codes with default $q$-sampling, but note that sparser sampling could make CLASS up to $2\times$ faster at unchanged precision
and shift its curves down by $\lg 2 \approx 0.3$.
\citet[in Appendix A]{howlettCMBPowerSpectrum2012} also found a precise three- to four-point scheme in CAMB,
and \citet{leeRapidAccurateNumerical2025} sped up CLASS using integral equations for massive neutrinos.

Despite the lack of approximation schemes, SymBoltz is fast thanks to
high-order implicit solvers,
fast generated functions for $\boldsymbol{f}$ and the analytical $\boldsymbol{J}$,
sparse matrix methods with a constant precomputed sparsity pattern,
and efficient sampling of massive neutrino momenta.
However, SymBoltz currently computes $C_l$ slower than CAMB and CLASS.
This computation is subject to more optimizations than $P(k)$, such as sampling of source functions $S(\tau,k)$, interpolation in $\tau$, $k$, and $l$, evaluation of spherical Bessel functions $j_l(x)$, and fast line-of-sight quadrature.
These aspects are not yet as polished in SymBoltz as in CAMB and CLASS, so a detailed performance analysis for $C_l$ (beyond Table \ref{tab:timings}) is not meaningful now.
Further work can add these optimizations to SymBoltz.
The purpose of the comparison made here is to show the viability of approximation-free perturbations, which marks a milestone in the design of SymBoltz.

\section{Future work}
\label{sec:futurework}

We hope SymBoltz continues to grow and that others can build on it.
Its symbolic capabilities can be extended with utilities for:
\begin{itemize}
\item separation into finer computational stages (see Sect. \ref{sec:splining});
\item gauge transformation of equations (e.g., Newtonian $\leftrightarrow$ synchronous gauge), as is already possible in CAMB;\footnote{\url{https://camb.readthedocs.io/en/latest/symbolic.html}}
\item perturbation theory utilities to derive initial conditions and approximation schemes for new models;
\item transformation of differential equations with respect to conformal time $\tau$ to another (one-to-one) independent variable (e.g., to get observables as direct functions of redshift $z$);
\item validation of dimensions in symbolic equations to catch modeling mistakes by tagging variables with physical units;
\item symbolic tensor algebra for work with modified gravity.
\end{itemize}
These tasks traditionally involve manual error-prone calculations, so such utilities could aid in the exploration of extended models.
In particular, we also aim to improve performance for differentiable runs to the level of standard runs.
An attractive milestone is fast reverse-mode gradients of scalar loss functions for gradient-based MCMCs.
We anticipate the symbolic structure and Jacobian generation to be useful for this,
as writing $\boldsymbol{u}(\tau,\boldsymbol{p})$ in the ODE \eqref{eq:ode} and differentiating it for $\boldsymbol{v}_i = \partial \boldsymbol{u} / \partial p_i$ gives
$\mathrm{d}\boldsymbol{v}_i/\mathrm{d}\tau = \boldsymbol{J} \boldsymbol{v}_i + \partial \boldsymbol{f} / \partial p_i$, which could be constructed analytically.

Future work could also generalize the code from scalar to vector and tensor perturbations, from flat to curved spacetimes, and from adiabatic to other initial conditions.
Some additional models for quintessence dark energy and Brans–Dicke gravity are in the works, and we invite others to add extended models.

We hope SymBoltz can foster growth of modular interoperating packages for cosmology in Julia.
For example, applications such as higher order perturbation theory (e.g., CLASS-PT by \citealt{chudaykinNonlinearPerturbationTheory2020} and PyBird by \citealt{damicoLimits$w$CDMEFTofLSS2021}), nonlinear boosting (e.g., HALOFIT by \citealt{takahashiRevisingHalofitModel2012} and HMCODE by \citealt{meadHMcode2020ImprovedModelling2021}), CMB lensing (e.g., CMBLensing.jl by \citealt{milleaBayesianDelensingDelight2020}), and nonlinear $N$-body simulations all rely on output from Boltzmann solvers.
Supporting automatic differentiation across programming languages is hard,
so an ecosystem of such packages in one language with support for automatic differentiation, such as Julia or Python with JAX, could create a modular and differentiable cosmological modeling toolbox.
A recent example of this is DISCO-DJ's coupling of differentiable Boltzmann and $N$-body codes for end-to-end modeling from linear perturbations to nonlinear structure formation \citep{hahnDISCODJDifferentiableEinsteinBoltzmann2024,listDISCODJIIDifferentiable2025}.

\section{Conclusion}
\label{sec:conclusion}

SymBoltz is a fresh Julia package for solving the linear Einstein–Boltzmann equations.
It features a symbolic-numeric interface where models are built from symbolic equations and compiled to efficient numerical code.
This lets users prototype new models with few lines of high-level code.
It relaxes all approximation schemes found in older codes by solving a single set of stiff equations at all times with implicit integrators.
This simplifies the equations and remains fast due to high-order implicit solvers that integrate efficient generated code for ODEs with analytical and sparse Jacobians.
It is differentiable, so we can get accurate derivatives of any output quantity with respect to any input parameter.
Output power spectra agree with CLASS to $0.1\%$ with standard precision, so SymBoltz can be used to fit current data.

\begin{figure}
    \centering
        \begin{tikzpicture}[
        pillar/.style={circle, draw, minimum size=2.4cm, align=center, font=\small},
        joiner/.style={midway, sloped, black, font=\small, above, yshift=+5pt},
        arrow/.style={-{Latex[width=4mm,length=4mm]}, line width=1mm, black},
    ]
    \node[pillar, fill=green!50] (sym) at (0:0cm) {Symbolic-\\numeric\\interface};
    \node[pillar, fill=blue!50] (app) at (240:5.7cm) {Approximation-\\freeness};
    \node[pillar, fill=red!50] (dif) at (300:5.7cm) {Differentiability};
    \draw[arrow] (sym.300) -- node[joiner] {Detect sparsity of $\boldsymbol{J}$} (dif.120);
    \draw[arrow] (dif.180) -- node[joiner] {Calculate $\boldsymbol{J}$ robustly} (app.0);
    \draw[arrow] (app.60) -- node[joiner] {Easy to specify model} (sym.240);
    \end{tikzpicture}
    \caption{%
        A symbolic-numeric interface, approximation-freeness, and differentiability are three main features of SymBoltz that form a synergy.
        Differentiability computes the ODE Jacobian $\boldsymbol{J}$ accurately and efficiently; which implicit solvers use to integrate stiff ODEs without approximations; which makes it easy to write equations in simple symbolic form; which provide the exact sparsity pattern of $\boldsymbol{J}$; in turn speeding up $\boldsymbol{J}$ and the solver.
        This loop creates a self-reinforcing design.
    }
    \label{fig:synergy}
\end{figure}

Other recent codes, such as PyCosmo, Bolt, and DISCO-EB, also incorporate some of these features.
SymBoltz combines them and reinforces its design with a synergy between them, as explained in Fig. \ref{fig:synergy}.
Together, these new codes offer a simpler alternative in high-level languages
that complements traditional codes such as CAMB and CLASS, which are historically highly tuned around approximations in low-level languages.
However, new codes need more work to reach the level of features and maturity of CAMB and CLASS that have grown over many years.

SymBoltz version 1.0.0 includes the flat Newtonian gauge metric, general relativity, cold dark matter, photons, baryons and RECFAST recombination, massless and massive neutrinos, the cosmological constant, and $w_0 w_a$ dark energy with scalar perturbations, distances, and matter and CMB spectra.
Forward-mode automatic differentiation works for single-run purposes such as Fisher forecasting.
Fast reverse-mode scalar loss gradients, particularly through the perturbations,
is a future milestone that could challenge emulators in gradient-based MCMC sampling.
No differentiable Boltzmann code is capable of this yet.

SymBoltz is publicly available\footnote{\url{https://github.com/hersle/SymBoltz.jl}} and easy to install.
Documentation is also available,\footnote{\url{https://hersle.github.io/SymBoltz.jl/}} and the code is tested with continuous integration to remain up-to-date (see Sect. \ref{sec:testing}).
All questions and contributions are welcome in the repository.
We hope SymBoltz proves useful and that it offers valuable ideas on the Boltzmann solver market.

\begin{acknowledgements}
I thank the Julia community for creating an extensive open ecosystem of scientific packages underlying SymBoltz.
I thank Aayush Sabharwal and Christopher Rackauckas for developing ModelingToolkit and OrdinaryDiffEq and answering questions.
I thank Hans A. Winther for helpful suggestions and feedback on the code and drafts of this paper.
I thank Julien Lesgourgues, Thomas Tram, and others for developing and thoroughly documenting CLASS.
I thank the anonymous referee for detailed and constructive comments that helped improve the quality of this paper.
This research was supported by the Research Council of Norway under project number 325113.
\end{acknowledgements}

\bibliographystyle{aa}
\bibliography{aa57450-25}

\appendix
\onecolumn

\section{List of equations and practical implementation details}
\label{sec:implementation}

This appendix summarizes the equations that define the standard \LCDM{} model in SymBoltz and comments on their practical implementation.
We hope this can be a useful reference for others.
The list mirrors the internal structure of SymBoltz with one component per subsection.
Unless otherwise stated, temperatures are in $\mathrm{K}$ and other variables are in units where \enquote{$G = c = H_0 = 1$.}
These units are chosen because $G$, $c$, and $H_0$ can be divided out from the Einstein equations as natural units (but this requires some conversion in the recombination equations that depend explicitly on $H_0$).
In other words, times are in units of $1/H_0$, distances in $c/H_0$, and masses in $c^3/H_0 G$.
When $G$, $c$, and $H_0$ appear explicitly in the equations, they are in SI units and used only to convert from SI units into the dimensionless units.
The equations closely follow the conventions in the seminal paper by \citet{maCosmologicalPerturbationTheory1995} and very closely match the source code of SymBoltz with Unicode characters.
It is far outside the scope of this paper to derive the equations and explain the meaning of every variable.

The independent variable is conformal time $\tau$, and all derivatives ${}^\prime = \mathrm{d} / \mathrm{d}\tau$ are with respect to it (in units of $1/H_0$).
This is perhaps the most common parameterization in the literature, and it is natural because most equations are autonomous with respect to $\tau$ (i.e., $\boldsymbol{f}(\boldsymbol{u},\tau) \rightarrow \boldsymbol{f}(\boldsymbol{u})$, except some multipole truncation schemes that have factors of $1/(k\tau)$, but these are not of physical origin).
By default, integration starts from the early time $\tau = \tau_i = 10^{-6}$ and terminates when the scale factor crosses $a = 1$ at the time $\tau = \tau_0$ today.

\subsection{Metric and spacetime \texorpdfstring{$(g)$}{(g)}}
\label{sec:metric}
SymBoltz is currently only formulated in the conformal Newtonian gauge, with this metric and related quantities:
\begin{align*}
    g_{0i} = g_{i0} = -a^2 (1+2\Psi) \delta_{0i}, \qquad
    g_{ij} = a^2 (1-2\Phi) \delta_{ij}, \qquad
    z = \frac{1}{a} - 1, \qquad
    \scrH = \frac{a^\prime}{a}, \qquad
    H = \frac{\scrH}{a} , \qquad
    \chi = \tau_0 - \tau .
\end{align*}
Here, $\scrH$ and $H$ are the conformal and cosmic Hubble factors (in units where $\scrH_0 = H_0 = 1$).
The scale factor $a$ is related to redshift $z$, and $\chi$ is the lookback time from today that appears in some integral solutions.
SymBoltz is currently restricted to a flat spacetime.

\subsection{General relativity \texorpdfstring{$(G)$}{(G)}}
\label{sec:gravity}

The gravitational theory of the \LCDM{} model is general relativity governed by the Einstein field equations $G_{\mu\nu} = 8\pi T_{\mu\nu}$.
By default, SymBoltz solves for the metric variables $a$, $\Phi$, and $\Psi$ with $(\mu,\nu) = (0,0)$ in the background (first Friedmann equation), and $(\mu,\nu) = \{(0,0), (i,j)\}$ in the perturbations:
\begin{align*}
a^\prime = \sqrt{\frac{8\pi}{3} \rho} \, a^{2}, \qquad
\Phi^\prime = - \scrH \Psi - \frac{k^{2}}{3 \scrH} \Phi - \frac{4\pi}{3} \frac{a^{2} }{\scrH} {\delta\rho} , \qquad
\Psi = - \Phi - 12\pi \bigg( \frac{a}{k} \bigg)^2 \Pi .
\end{align*}
We note that it is also possible to evolve other redundant combinations of the Einstein equations, such as the acceleration equation.
The equations are coupled to total densities $\rho$ and $\delta\rho$, pressures $P$ and $\delta P$, and anisotropic stress $\Pi$ for whichever set of species $s$ that are present in the cosmological model:
\begin{align*}
\rho = \sum_s \rho_s, \qquad
P = \sum_s P_s, \qquad
\delta\rho = \sum_s \delta\rho_s = \sum_s \delta_s \rho_s , \qquad
\delta P = \sum_s \delta P_s = \sum_s \delta_s \rho_s c_{s,s}^2, \qquad
\Pi = \sum_s \Pi_s = \sum_s (\rho_s + P_s) \sigma_s .
\label{eq:gravityspecies}
\end{align*}

We emphasize that the gravity component is completely unaware of all particle species and makes no assumptions about them.
It only reacts to total stress-energy components.
All species must therefore define $\rho$, $P$, $\delta \rho$, $\delta P$, and $\Pi$ explicitly, even if they are zero.
This requirement is somewhat pedantic, but helps isolate components from each other for greater reuse when composing models.

The scale factor $a$ is initialized as the nonlinear solution of the Friedmann equation constrained to $\scrH = 1/\tau$ (motivated by its radiation-dominated solution $a = \sqrt{\Omega_{r0}} \, \tau$).
The constraint potential is initialized to $\Psi = 20C/(15+4f_\nu)$ with the (arbitrary) integration constant $C=1/2$ and initial energy density fraction $f_\nu = (\rho_\nu+\rho_h) / (\rho_\nu+\rho_h+\rho_\gamma)$ of all (massless and massive) neutrino species relative to all species that are radiation-like at early times.
The evolved potential $\Phi$ is initialized accordingly from the constraint equation (the result is close to $\Phi = (1 + 2 f_\nu / 5) \Psi$, but providing both $\Phi$ and $\Psi$ explicitly leads to overdetermined initialization equations that violate the constraint equation for $\Psi$).
The next sections present the \LCDM{} part of the \enquote{library of species} that are available in SymBoltz.

\subsection{Cold dark matter \texorpdfstring{$(c)$}{(c)}}
\label{sec:cdm}
Cold dark matter is a nonrelativistic and noninteracting species that follows very simple equations:
\begin{align*}
w = 0 , \qquad
{c_s^2} = w , \qquad
P = 0 , \qquad
\rho = \frac{\rho_0}{a^{3}} , \qquad
\delta^\prime = - \theta + 3 \Phi^\prime , \qquad
\theta^\prime = - \scrH \theta + k^{2} \Psi , \qquad
u = \frac{\theta}{k} , \qquad
\sigma = 0 .
\end{align*}
Initial conditions are adiabatic with $\delta/(1+w) = -3 \Psi / 2$ and $\theta = k^2 \tau \Psi / 2$.
The species is parameterized by the reduced density $\Omega_0 = \frac{8\pi}{3}\rho_0$ today.

\subsection{Baryons \texorpdfstring{$(b)$}{(b)}}
\label{sec:baryons}
Baryons are also nonrelativistic, but interact with photons through Compton scattering and are subject to recombination physics.
This significantly complicates their behavior.
SymBoltz currently implements equations from RECFAST\footnote{\url{https://www.astro.ubc.ca/people/scott/recfast.html}} version 1.5.2 \citep{seagerNewCalculationRecombination1999,wongHowWellWe2008,scottMatterTemperatureCosmological2009}:
\begin{align*}
&
w = 0 , \qquad
P = 0 , \qquad
\rho = \frac{\rho_0}{a^{3}} , \qquad
{f_\He} = \frac{{Y_\He}}{\frac{m_\He}{m_\Hy} \big( 1 - {Y_\He} \big)} , \qquad
n_\Hy = \frac{(1-Y_\He) \rho}{m_\Hy}, \qquad
n_\He = f_\He n_\Hy , \qquad
{c_s^2} = \frac{k_B}{\mu c^2} \bigg( {T_b} - \frac{T_b^\prime}{3 \scrH} \bigg) , \\
&
\beta = \frac{1}{k_B T_b} , \qquad
T_b^\prime = - 2 \scrH {T_b} - \frac{a}{H_0} \frac{8}{3} \frac{ T_\gamma^{4} {X_\el}}{1 + {f_\He} + {X_\el}} \big(T_b - T_\gamma\big) , \qquad
{\mu} = \frac{m_\Hy}{1 + \big(\frac{m_\Hy}{m_\He}-1\big) {Y_\He} + \big( 1 - {Y_\He} \big) {X_\el}} , \qquad
\kappa^\prime = -\frac{a}{H_0} n_\el \sigma_T c , \\
&
{v} = - \kappa^\prime e^{-{\kappa}} , \qquad
n_\el = X_\el n_\Hy, \qquad
{X_\el} = {X_\Hy^+} + {X_\He^{++}} + {f_\He} {X_\He^+} + {X_\el^\reone} + {X_\el^\retwo} , \qquad
X_\Hy^{+\prime} = -\frac{a}{H_0} {C_\Hy} \Big( {\alpha}_\Hy n_\el {X_\Hy^+} - {\beta_\Hy} e^{ - {\beta_b} E_\Hy^{2s,1s} } \big( 1 - {X_\Hy^+} \big) \Big) , \\
&
{X_\Hesin^{+\prime}} = -\frac{a}{H_0} C_\Hesin \Big( \alpha_\Hesin n_\el X_\He^+ - \beta_\Hesin e^{-\beta_b E_\Hesin^{2s,1s} } \big( 1 - X_\He^+ \big)  \Big) , \qquad
X_\Hetri^{+\prime} = -\frac{a}{H_0} C_\Hetri \Big( n_\el {\alpha_\Hetri} {X_\He^+} - 3 {{\beta}_\Hetri} e^{ - {\beta_b} E_\Hetri^{2s,1s} } \big( 1 - {X_\He^+} \big) \Big) , \\
&
{X_\He^{+\prime}} = {X_\Hesin^{+\prime}} + {X_\Hetri^{+\prime}} , \qquad
{X_\He^{++}} = \frac{2 {f_\He} {R_\He^+}}{\bigg( 1 + f_\He + R_\He^+ \bigg) \bigg( 1 + \sqrt{1 + \frac{4 {f_\He} {R_\He^+}}{( 1 + f_\He + R_\He^+)^{2}}} \bigg)} , \qquad
R_{\He^+} = \frac{\exp\big({-\beta E_{\He^+}^{\infty,1s}}\big)}{n_\Hy \lambda_\el^3} , \qquad
\lambda_\el = \frac{h}{\sqrt{2\pi m_\el/\beta}} , \\
&
{X_\el^\reone} = \frac{1 + f_\He}{2} + \frac{ 1 + f_\He }{2} \tanh\bigg( \frac43 \frac{( 1 + z^\reone )^{3/2} - ( 1 + z )^{3/2}}{ ( 1 + z^\reone )^{1/2}} \bigg) , \qquad
{X_\el^\retwo} = \frac{f_\He}{2} + \frac{f_\He}{2} \tanh\bigg( \frac43 \frac{( 1 + {z^\retwo} )^{3/2} - ( 1 + z )^{3/2}}{ ( 1 + {z^\retwo} )^{1/2}} \bigg) , \\
&
\delta^\prime = - \theta - 3 \scrH c_s^2 \delta + 3 \Phi^\prime , \qquad
\theta^\prime = - \scrH \theta + k^{2} {c_s^2} \delta + k^{2} \Psi - \frac43 \kappa^\prime \frac{\rho_\gamma}{\rho_b} \big(\theta_\gamma-\theta_b\big) , \qquad
u = \frac{\theta}{k} , \qquad
\sigma = 0 .
\end{align*}
Transition rates and coefficients related to recombination of Hydrogen include fitting functions that emulate the results of more accurate and expensive computations (here $\ln(a)$ is the logarithm of the scale factor, while $(Fa)$ is an unrelated fudge factor):
\begin{align*}
&
\alpha_\Hy = 10^{-19} (Fa) \frac{\Big( \frac{T_b}{T_0} \Big)^b }{ 1 + c \Big(\frac{T_b}{T_0}\Big)^d }, \qquad
\beta_\Hy = \frac{\alpha_\Hy}{\lambda_\el^3} \exp\big({-\beta E_\Hy^{\infty,2s}}\big), \\
&
K_\Hy = \Bigg( 1 + A_1 \exp\Bigg({-\bigg(\frac{\ln(a)-\ln(a_1)}{w_1}\bigg)^2}\Bigg) + A_2 \exp\Bigg({-\bigg(\frac{\ln(a)-\ln(a_2)}{w_2}\bigg)^2}\Bigg) \Bigg) \frac{\big(\lambda_\Hy^{2s,1s}\big)^3}{8\pi H}, \qquad
C_\Hy = \frac{1 + K_\Hy \Lambda_\Hy n_\Hy (1-X_\Hy^+) }{ 1 + K_\Hy (\Lambda_\Hy+\beta_\Hy) n_\Hy (1-X_\Hy^+) }.
\end{align*}
Helium rates and coefficients are even more complicated. First, Helium includes contributions from singlet states ($\Hesin$):
\begin{align*}
&
\alpha_\Hesin = \frac{q_1 }{ \sqrt{\frac{T_b}{T_2}} \Big(1+\sqrt{\frac{T_b}{T_2}}\Big)^{1-p_1} \Big(1+\sqrt{\frac{T_b}{T_1}}\Big)^{1+p_1} }, \qquad
\beta_\Hesin = 4 \frac{\alpha_\Hesin }{ \lambda_\el^3 } \exp\big({-\beta E_\Hesin^{\infty,2s}}\big), \qquad
K_\Hesin = \frac{1}{K_{\He_1^0}^{-1} + K_{\He_1^1}^{-1} + K_{\He_1^2}^{-1}}, \\
&
K_{\He_1^0}^{-1} = \frac{8 \pi H }{ \big(\lambda_\Hesin^{2p,1s}\big)^3 }, \qquad
K_{\He_1^1}^{-1} = -\exp(-\tau_\Hesin) K_{\He_1^0}^{-1}, \qquad
K_{\He_1^2}^{-1} = \frac{A_{2p_1} }{ 3 \big(1+0.36 \, \gamma_{2p_1}^{0.86}\big) n_\He (1-X_\He^+) }, \qquad
\tau_\Hesin = \frac{ 3 A_{2p_1} n_\He (1-X_\He^+) }{ K_{\He_1^0}^{-1} }, \\
&
\gamma_{2p_1} = \frac{ 3 A_{2p_1} f_\He c^2 (1-X_\He^+) }{ 8\pi \sigma_{\Hesin} \sqrt{\frac{2 \pi}{\beta m_\He c^2}} \big(f_\Hesin^{2p,1s}\big)^3 (1-X_\Hy^+) }, \qquad
C_\Hesin = \frac{\exp\big({-\beta E_\Hesin^{2p,2s}}\big) + K_\Hesin \Lambda_\Hesin n_\He (1-X_\He^+) }{ \exp\big({-\beta E_\Hesin^{2p,2s}}\big) + K_\Hesin (\Lambda_\Hesin+\beta_\Hesin) n_\He (1-X_\He^+) }.
\end{align*}
Second, Helium also includes contributions from triplet states ($\Hetri$):
\begin{align*}
&
\alpha_\Hetri = \frac{q_3 }{ \sqrt{\frac{T_b}{T_2}} \Big(1+\sqrt{\frac{T_b}{T_2}}\Big)^{1-p_3} \Big(1+\sqrt{\frac{T_b}{T_1}}\Big)^{1+p_3} }, \qquad
\beta_\Hetri = \frac43 \frac{ \alpha_\Hetri }{ \lambda_\el^3 } \exp\big({-\beta E_\Hetri^{\infty,2s}}\big), \qquad
\tau_\Hetri = \frac{ 3 A_{2p_3} n_\He (1-X_\He^+) \big(\lambda_\Hetri^{2p,1s}\big)^3 }{ 8\pi H }, \\
&
\gamma_{2p_3} = \frac{ 3 A_{2p_3} f_\He c^2 (1-X_\He^+) }{ 8\pi \sigma_{\Hetri} \sqrt{\frac{2 \pi}{\beta m_\He c^2}} \big(f_\Hetri^{2p,1s}\big)^3 (1-X_\Hy^+) }, \qquad
C_\Hetri = \frac{ A_{2p_3} \bigg( \frac{1 - \exp({-\tau_\Hetri}) }{ \tau_\Hetri } + \frac{1}{3\big(1+0.66\,\gamma_{2p_3}^{0.9}\big)} \bigg) \exp\big({-\beta E_\Hetri^{2p,2s}}\big) }{ A_{2p_3} \bigg( \frac{1 - \exp({-\tau_\Hetri}) }{ \tau_\Hetri } + \frac{1}{3\big(1+0.66\,\gamma_{2p_3}^{0.9}\big)} \bigg) \exp\big({-\beta E_\Hetri^{2p,2s}}\big) + \beta_\Hetri } .
\end{align*}
Every variable that does not occur on the left side of an equation is either a constant or a parameter given in the code.
This includes $Y_\He$, fudge factors, and wavenumbers, frequencies, and energies for atomic transitions.
Some important variables defined above are the baryon temperature $T_b$, photon temperature $T_\gamma$, mean molecular weight $\mu$, baryon sound speed $c_s^2$, optical depth $\kappa$, visibility function $v$, and the free electron fraction $X_\el$ (conventionally relative to Hydrogen, so $X_\el > 1$ in presence of Helium).
We refer to the code and RECFAST references cited above for more details.

Unlike other RECFAST implementations, SymBoltz does not approximate the stiff Peebles equations at early times by Saha approximations (although $X_\He^{++}$ is given by a Saha equation at all times).
This is not necessary with a good implicit ODE solver.
SymBoltz sets $C_\Hy = C_\Hesin = 1$ when $X_\el \gtrsim 0.99$ to avoid numerical instability at early times.
Atomic calculations are done in SI units and converted to SymBoltz' dimensionless units by factors of $H_0$ in SI units.
The differential equation for $T_b^\prime$ is very stiff and sensitive to $T_b-T_\gamma$, but $T_b \approx T_\gamma$ in the early universe, so we rewrite it to a more stable differential equation for $\Delta T^\prime = T_b^\prime - T_\gamma^\prime$ instead, initialize $\Delta T = 0$, and observe $T_b = \Delta T + T_\gamma$.
The optical depth $\kappa(\tau) = \int_{\tau_0}^\tau \kappa^\prime(\tau^\prime) \mathrm{d}\tau^\prime$ is really a line-of-sight integral into the past,
but we solve it together with the background ODEs by initializing $\kappa(\tau_i) = 0$ to an arbitrary value, integrating the differential equation for $\kappa^\prime$, and subtracting the final value of $\kappa(\tau_0)$ (i.e., $\int_{\tau_0}^{\tau} = \int_{\tau_0}^{\tau_i} + \int_{\tau_i}^{\tau} = \int_{\tau_i}^{\tau} - \int_{\tau_i}^{\tau_0}$).
There is no tight-coupling approximation.

Initial conditions are fully ionized fractions $X_\Hy^+ = X_\He^+ = 1$, temperatures $T_b = T_\gamma$ ($\Delta T = 0$) in equilibrium, the arbitrary $\kappa = 0$, and adiabatic perturbations $\delta/(1+w) = -3 \Psi / 2$ and $\theta = k^2 \tau \Psi / 2$.
The baryon species is parameterized by the reduced density $\Omega_0 = \frac{8\pi}{3}\rho_0$ today, and the primordial Helium mass fraction $Y_\He$.

SymBoltz solves thermodynamics equations together with the background equations, while some other codes treat these as separate stages.
There is no meaningful performance improvement from doing this, as the size of the background (and thermodynamics) ODEs is so small.
Solving these stages together creates a clear distinction between the background with all zeroth-order equations of motion, and the perturbations with all first-order equations.
It also makes it possible to create exotic models where the thermodynamics affect the background, for example.

We note that RECFAST uses fitting functions to emulate the results of more physically accurate and expensive simulations.
These are tuned to work for the \LCDM{} model.
SymBoltz would therefore benefit from including more realistic recombination models for safer use with modified models.

\subsection{Photons \texorpdfstring{$(\gamma)$}{(γ)}}
\label{sec:photons}
Photons are massless and therefore ultrarelativistic.
Unlike nonrelativistic particles, we must account for the direction $\cos\theta = \boldsymbol{p} \cdot \boldsymbol{k} \,/\, \lvert\boldsymbol{p}\rvert \cdot \lvert\boldsymbol{k}\rvert$ of their momenta $\boldsymbol{p}$ relative to the Fourier wavenumber $\boldsymbol{k}$.
This results in a theoretically infinite hierarchy of equations for Legendre multipoles $l$, which in practice must be truncated at some maximum multipole $\lmax$:
\begin{align*}
&
T = \frac{{T_0}}{a} , \qquad
w = \frac{1}{3} , \qquad
{c_s^2} = w , \qquad
P = \frac{\rho}{3} , \qquad
\rho = \frac{\rho_0}{a^{4}} , \qquad
\delta = F_0 , \qquad
\theta = \frac{3}{4} k F_{1} , \qquad
u = \frac{\theta}{k}, \qquad
\sigma = \frac{F_2}{2} , \\
&
F_0^\prime = - k F_1 + 4 \Phi^\prime , \qquad
F_1^\prime = \frac{k}{3} \big( F_0 - 2 F_2 + 4 \Psi \big) + \frac{4}{3} \frac{\kappa^\prime}{k} \big( \theta_\gamma - {\theta_b} \big), \\
&
F_l^\prime = \frac{k}{2l+1} \big( l F_{l-1} - (l+1) F_{l+1} \big) + F_{l} {\kappa^\prime} - \delta_{l,2}\frac{\kappa^\prime}{10} \Pi , \qquad
F_{\lmax}^\prime = k F_{\lmax-1} - \frac{\lmax+1}{\tau} F_{\lmax} + {\kappa^\prime} F_{\lmax} , \\
&
G_0^\prime =  - k G_{1} + {\kappa^\prime} {G_0} - \frac{\kappa^\prime}{2} \Pi , \qquad
G_{l}^\prime = \frac{k}{2l+1} \big( l G_{l-1} - (l+1) G_{l+1} \big) + \kappa^\prime G_{l} - \delta_{l,2} \frac{{\kappa^\prime}}{10} \Pi  ,  \\
&
G_{\lmax}^\prime = k G_{\lmax-1} - \frac{\lmax+1}{\tau} G_{\lmax} + {\kappa^\prime} G_{\lmax} , \qquad
\Pi = F_{2} + G_0 + G_{2} , \qquad
{{\Theta}_l} = \frac{F_l}{4} .
\end{align*}
The equations for $F_l^\prime$ and $G_l^\prime$ hold for $2 \leq l < \lmax$.
There are no tight-coupling, radiation-streaming or ultrarelativistic fluid approximations.
Initial conditions are adiabatic with $F_0 = -2\Psi$ (i.e., $\delta/(1+w) = -\frac32 \Psi$), $F_1 = \frac23 k\tau \Psi$ (i.e., $\theta = \frac12 k^2 \tau \Psi$), $F_2 = -\frac{8}{15} \frac{k}{\kappa^\prime} F_1$, $G_0 = \frac{5}{16} F_2$, $G_1 = -\frac{1}{16} \frac{k}{\kappa^\prime} F_2$, $G_2 = \frac{1}{16} F_2$, $F_l = -\frac{l}{2l+1} \frac{k}{\kappa^\prime} F_{l-1}$, and $G_l = -\frac{l}{2l+1} \frac{k}{\kappa^\prime} G_{l-1}$ for $3 \leq l \leq \lmax$.
The species is parameterized by its temperature $T_0$ today, which in turn sets the density parameters $\Omega_0 = \frac{\pi^2}{15} \frac{(k_B T_0)^4}{(\hbar c)^3} \frac{8\pi G}{3H_0^2}$ and $\rho_0 = \frac{8\pi}{3}\Omega_0$ today.

\subsection{Massless neutrinos \texorpdfstring{$(\nu)$}{(ν)}}
\label{sec:masslessneutrinos}
Massless neutrinos behave similarly to photons, but decouple from interactions with other species in the very early universe.
We must only account for this interaction in initial conditions, while their evolution equations are a simpler case of the photon equations:
\begin{align*}
&
T = \frac{T_0}{a} , \qquad
w = \frac{1}{3} , \qquad
{c_s^2} = \frac{1}{3} , \qquad
P = \frac{\rho}{3} , \qquad
\rho = \frac{\rho_0}{a^{4}} , \qquad
\delta = {F_0} , \qquad
\theta = \frac{3}{4} k F_{1} , \qquad
\sigma = \frac{F_2}{2} , \\
&
F_0^\prime = - k F_{1} + 4 \Phi^\prime , \qquad
F_{1}^\prime = \frac{k}{3} \big( {F_0}  - 2 F_{2} + 4 \Psi \big) , \qquad
F_{l}^\prime = \frac{k}{2l+1} \big( l F_{l-1} - (l+1) F_{l+1} \big) , \qquad
F_{\lmax}^\prime = k F_{\lmax-1} - \frac{\lmax+1}{\tau} F_{\lmax} .
\end{align*}
The equations for $F_l^\prime$ hold for $2 \leq l < \lmax$.
There is no ultrarelativistic fluid approximation.
Initial conditions are adiabatic with $F_0 = -2 \Psi$ (i.e., $\delta/(1+w) = -\frac32 \Psi$), $F_1 = \frac23 k \tau \Psi$ (i.e., $\theta = \frac12 k^2 \tau \Psi$), $F_2 = \frac{2}{15} (k\tau)^2 \Psi$, and $F_l = \frac{l}{2l+1} k\tau F_{l-1}$.
The species is parameterized by the effective number $N_\text{eff}$, the reduced density $\Omega_0 = \frac{8\pi}{3} \rho_0$ today, and temperature $T_0$ today.
If photons are present, they default to $T_{\nu 0} = \big(\frac{4}{11}\big)^{1/3} T_{\gamma 0}$ and $\Omega_{\nu 0} = N_\text{eff} \frac78 \big(\frac{4}{11}\big)^{4/3} \Omega_{\gamma 0}$.

\subsection{Massive neutrinos \texorpdfstring{$(h)$}{(h)}}
\label{sec:massiveneutrinos}

Massive neutrinos are the most complicated species in the \LCDM{} model (alongside baryon recombination).
In essence, the species we have looked at so far have Boltzmann equations where the momenta of their distribution function can be integrated out analytically in nonrelativistic or ultrarelativistic limits.
This means that their stress-energy components are linked by trivial equations of state and sound speeds, for example, and their effect can be parameterized by a simple density parameter $\Omega_{0}$.

On the other hand, massive neutrinos have intermediate masses that fall between the nonrelativistic and ultrarelativistic limits.
Integrals over their distribution function must be computed numerically.
This is very expensive if done naively, and it is extremely important to choose a quadrature scheme that samples as few momenta as possible.
Fortunately, the momentum integrals have a structure that can be exploited: they are all in the weighted form $I[g(x)] = \int_0^\infty \mathrm{d}x \, x^2 f(x) g(x)$, where $f(x) = 1/(e^x+1)$ is the equilibrium distribution function and $g(x)$ is an arbitrary function of the dimensionless momentum $x = pc/k_B T$ (see \citealt{maCosmologicalPerturbationTheory1995} for more details).
We can generally approximate $I[g(x)] \approx \sum_i W_i \, g(x_i)$ with a weighted quadrature scheme with points $x_i$ and weights $W_i$ (more on this after the equations).
In other words, the integral operator $\int_0^\infty \mathrm{d} x \, x^2 f(x)$ is effectively replaced by the discrete summation operator $\sum_i \! W_i$ for some weights $W_i$.
On top of this, perturbations are also expanded in Legendre multipoles $l$ up to a cutoff $\lmax$:
\begin{align*}
&
T = \frac{{T_0}}{a} , \qquad
x = \frac{pc}{k_B T}, \qquad
y = \frac{mc^2}{k_B T}, \qquad
E_i = \sqrt{x_i^2 + y^{2}} , \qquad
f = \frac{1}{1 + e^x} , \qquad
\dlnfdlnx = -\frac{x}{1 + e^{-x}}, \\
&
I_\rho = \sum_i \! W_i E_i , \qquad
I_P = \sum_i \! W_i \frac{x_i^2}{E_i}, \qquad
\rho = \frac{N}{\pi^2} \frac{(k_B T)^4}{(\hbar c)^3} \frac{G}{(H_0 c)^2} I_\rho , \qquad
P = \frac{N}{3\pi^2} \frac{(k_B T)^4}{(\hbar c)^3} \frac{G}{(H_0 c)^2} I_P , \qquad
w = \frac{P}{\rho} , \\
&
\psi_{i,0}^\prime = -k \frac{x_i}{E_i} \psi_{i,1} - \Phi^\prime \left( \dlnfdlnx \right)_i , \qquad
\psi_{i,1}^\prime = \frac{k}{3} \frac{x_i}{E_i} \big( \psi_{i,0} - 2 \psi_{i,2} \big) - \frac{k}{3} \frac{E_i}{x_i} \Psi \left( \dlnfdlnx \right)_i , \\
&
\psi_{i,l}^\prime = \frac{k}{2l+1} \frac{x_i}{E_i} \big( l \psi_{i,l-1} - (l+1) \psi_{i,l+1} \big) , \qquad
\psi_{i,\lmax+1} = \frac{2\lmax+1}{k \tau} \frac{E_i}{x_i} \psi_{i,\lmax} - \psi_{i,\lmax-1} , \\
&
I_{0} = \sum_i \! W_i E_i \psi_{i,0}, \qquad
I_{1} = \sum_i \! W_i x_i \psi_{i,1}, \qquad
I_{2} = \sum_i \! W_i \frac{x_i^2}{E_i} \psi_{i,2} , \qquad
\delta = \frac{I_{0}}{I_\rho} , \qquad
\sigma = \frac{ 2 I_2}{3 I_\rho + I_P} , \qquad
u = \frac{3 I_1}{3I_\rho + I_P} , \qquad
\theta = k u .
\end{align*}
Initial conditions are $\psi_{i,0} = -\frac{1}{4} (-2 \Psi) \big(\dlnfdlnx\big)_i$, $\psi_{i,1} = -\frac{1}{3} \frac{E_i}{x_i} \frac{1}{2} k\tau \Psi \big(\dlnfdlnx\big)_i$, $\psi_{i,2} = -\frac{1}{2} \frac{1}{15} (k\tau)^2 \Psi \big(\dlnfdlnx\big)_i$, and $\psi_{i,l} = 0$.
When integrated over momenta, they are equivalent to adiabatic $\delta/(1+w)$, $\theta$, and $\sigma$, similarly to massless neutrinos.
Free parameters are the temperature $T_0$ today, single neutrino mass $m$, and degeneracy factor $N = \sum_{i=1}^N m_i / m$ for describing multiple neutrinos with the same mass.
They default to $N = 3$, and $T_{h0} = \big(\frac{4}{11}\big)^{1/3} T_{\gamma 0}$ if photons are present, as for massless neutrinos.

Here, the equations for $\psi_{i,l}^\prime$ hold for $2 \leq l \leq \lmax$, and the $i$-indexed expressions $g_i = g(x_i)$ are evaluated with the momentum quadrature point $x = x_i$.
The reduction to dimensionless momenta $x = pc/k_B T$ (the argument of $\exp$ in $f$) is deliberate because it makes numerics more well-defined and the quadrature scheme independent of $m$ and all other cosmological parameters.

SymBoltz automatically computes momentum bins $x_i$ and quadrature weights $W_i$ with $N$-point Gaussian quadrature.
First, by default, the following substitution is applied to the momentum integral:
\begin{equation*}
    \int_0^\infty \mathrm{d}x \, x^2 f(x) g(x) = \int_{u(0)}^{u(\infty)} \mathrm{d}u \, x^\prime(u) x(u)^2 f(x(u)) g(x(u))
    \quad \text{with} \quad
    u(x) = \frac{1}{1+\frac{x}{L}} .
\end{equation*}
This substitution achieves two things: the scaling $x/L$ brings the dominant integral contributions well within $x/L \ll 1$ if $L$ is chosen to be a characteristic decay \enquote{length} of the distribution function, and the rational part $1/(1+x/L)$ maps the infinite domain $x \in (0, \infty)$ to the finite domain $u \in (0, 1)$, which can be integrated numerically.
The substituted integrand is then passed to an adaptive algorithm in QuadGK.jl\footnote{\url{https://github.com/JuliaMath/QuadGK.jl}} that computes quadrature points $u_i$ and weights $W_i$ by performing weighted integrals against several test functions $g(x)$.
Finally, the corresponding momenta $x_i = x(u_i)$ are returned along with the weights $W_i$, from which we can approximate the integral $I[g(x)] \approx \sum_i W_i g(x_i)$ against any $g(x)$.

We test this numerical quadrature scheme against the analytical result $I[x^{n-2}] = \int_0^\infty \mathrm{d}x \, x^n / (e^x+1) = (1-2^{-n}) \zeta(n+1) \Gamma(n+1)$ for $2 \leq n \leq 8$.
We assume this is a reasonable test for the integrals encountered in the equations above. 
The agreement is excellent with $L = 100$, which yields relative errors below $10^{-6+n-N}$ for all $2 \leq n \leq 8$ and $1 \leq N \leq 5$.
SymBoltz defaults to $N=4$ momenta, for which this relative error is less than $10^{-6}$ for $n \leq 4$, for example.
It also agrees well with CLASS using default settings.

We note that this momentum quadrature strategy is generic with respect to the distribution function $f(x)$ and substitution $u(x)$, so it can easily be modified for other particle species whose distribution function cannot be integrated out.

CAMB\footnote{\url{https://cosmologist.info/notes/CAMB.pdf}} and CLASS \citep{lesgourguesCosmicLinearAnisotropy2011a} apply similar weighted quadrature strategies.
They also get away with only a handful of sampled momenta, but the precise details of the computation differ slightly.
For reference, here are points and weights computed by SymBoltz for $1 \leq N \leq 8$ momenta:
\begin{center}
\begin{tabular}{  r r r r r r r r r  } 
\toprule
$N$ & $x_1$ & $x_2$ & $x_3$ & $x_4$ & $x_5$ & $x_6$ & $x_7$ & $x_8$ \\
\midrule
1 & 3.12273 & & & & & & & \\
2 & 2.07807 & 5.94834 & & & & & & \\
3 & 1.56110 & 4.22902 & 8.86258 & & & & & \\
4 & 1.24461 & 3.30909 & 6.57536 & 11.80351 & & & & \\
5 & 1.02955 & 2.71805 & 5.27853 & 9.03363 & 14.74043 & & & \\
6 & 0.87373 & 2.30142 & 4.41479 & 7.39595 & 11.55088 & 17.65818 & & \\
7 & 0.75572 & 1.99028 & 3.79064 & 6.27323 & 9.60568 & 14.09716 & 20.54878 & \\
8 & 0.66337 & 1.74848 & 3.31580 & 5.44442 & 8.24194 & 11.87257 & 16.65452 & 23.40806 \\
\midrule
$N$ & $W_1$ & $W_2$ & $W_3$ & $W_4$ & $W_5$ & $W_6$ & $W_7$ & $W_8$ \\
\midrule
1 & 1.80309 & & & & & & & \\
2 & 1.30306 & 0.50002 & & & & & & \\
3 & 0.84813 & 0.88596 & 0.06899 & & & & & \\
4 & 0.55272 & 0.99943 & 0.24384 & 0.00709 & & & & \\
5 & 0.36868 & 0.95311 & 0.43658 & 0.04409 & 0.00063 & & & \\
6 & 0.25275 & 0.84165 & 0.58496 & 0.11736 & 0.00632 & 0.00005 & & \\
7 & 0.17792 & 0.71541 & 0.67227 & 0.21284 & 0.02386 & 0.00079 & 0.00000 & \\
8 & 0.12832 & 0.59663 & 0.70569 & 0.31144 & 0.05685 & 0.00406 & 0.00009 & 0.00000 \\
\bottomrule
\end{tabular}
\end{center}

\subsection{Cosmological constant \texorpdfstring{$(\Lambda)$}{(Λ)}}
\label{sec:cc}

The cosmological constant is equivalent to a very simple species without perturbations:
\begin{align*}
    w = -1 , \qquad
    \rho = \rho_0, \qquad
    P = -\rho , \qquad
    \delta = 0 , \qquad
    \theta = 0 , \qquad
    \sigma = 0 , \qquad
    u = 0 .
\end{align*}
It is parameterized by the reduced density $\Omega_0 = \frac{8\pi}{3} \rho_0$ today.
If all species $s$ have a $\Omega_{s0}$ parameter and general relativity is the theory of gravity, it is set to $\Omega_{\Lambda 0} = 1 - \sum_{s \neq \Lambda} \Omega_{s0}$.
This constraint comes from the first Friedmann equation today.

\subsection{Primordial power spectrum \texorpdfstring{($I$)}{(I)}}
\label{sec:primordial}

SymBoltz computes the inflationary primordial power spectrum parameterized by the amplitude $A_s$ and tilt $n_s$:
\begin{align*}
    P_0(k) = \frac{2\pi^2}{k^3} A_s \bigg(\frac{k}{k_\text{p}}\bigg)^{n_s-1}.
\end{align*}

\subsection{Matter power spectrum}
\label{sec:matter}

SymBoltz computes the matter power spectrum for some desired set of species $s$, which are presumably matter-like at late times (e.g., $s \in \{c,b,h\}$):
\begin{align*}
    P(k,\tau) = P_0(k) \big|\Delta(\tau,k)\big|^2
    \quad \text{with} \quad
    \Delta = \delta + \frac{3 \scrH}{k^2} \theta = \frac{\sum_s \delta\rho_s}{\sum_s \rho_s} + \frac{3 \scrH}{k^2} \frac{\sum_s (\rho_s+P_s)\theta_s}{\sum_s (\rho_s+P_s)} .
\end{align*}
Here, $\Delta$ is the total gauge-independent overdensity with total $\delta$ and $\theta$ computed by summing the components of the energy-momentum tensor that are additive.

\subsection{CMB power spectrum and line-of-sight integration}

SymBoltz finds photon temperature and polarization multipoles today for any $l$ by computing line-of-sight integrals:
\label{sec:cmb}
\begin{align*}
    &
    \Theta_l^\mathrm{T}\big(\tau_0,k\big) = \int_{\tau_i}^{\tau_0} S_T(\tau,k) j_l\big(k(\tau_0-\tau)\big) \mathrm{d}\tau \quad \text{with} \quad
    S_T = v \bigg( \frac{\delta_\gamma}{4} + \Psi + \frac{\Pi_\gamma}{16} \bigg) + e^{-\kappa} \big( \Psi + \Phi \big)^\prime + \frac{\big(v u_b\big)^\prime}{k} + \frac{3}{16k^2}\big(v \Pi_\gamma\big)^{\prime\prime}, \\
    &
    \Theta_l^\mathrm{E}\big(\tau_0,k\big) = \sqrt{\frac{(l+2)!}{(l-2)!}} \int_{\tau_i}^{\tau_0} S_E\big(\tau,k\big) \frac{j_l\big(k(\tau_0-\tau)\big)}{\big(k(\tau_0-\tau)\big)^2} \mathrm{d}\tau \quad \text{with} \quad 
    S_E = \frac{3}{16} v \Pi_\gamma .
\end{align*}
As first suggested by \citet{seljakLineSightApproach1996}, this approach enables cheap computation for any $l$ after integrating the perturbation ODEs with only a few $l \leq \lmax$.
This drastically speeds up the computation over including all $l$ in an enormous set of coupled perturbation ODEs.
SymBoltz performs the integrals with the trapezoid method using around $1250$ uniformly sampled $\tau$, by default.
Here, $j_l$ are the spherical Bessel functions of the first kind.
SymBoltz is not yet generalized to curved spacetimes, where they are replaced by hyperspherical functions.
The cross-correlated angular spectrum between $\mathrm{A},\mathrm{B} \in \{\mathrm{T},\mathrm{E}\}$ is then computed from
\begin{equation*}
    C_l^\mathrm{AB} = \frac{2\pi}{l(l+1)} D_l^\mathrm{AB} = \frac{2}{\pi} \int_0^\infty \mathrm{d}k k^2 P_0(k) \, \Theta_l^A(\tau_0,k) \, \Theta_l^B(\tau_0,k) .
\end{equation*}
This integral is also performed with the trapezoid method.
The point $(k, \Theta) = (0, 0)$ is included manually, for which the numerical solution to the perturbation ODEs is ill-defined.
By default, the $\Theta_l$ are sampled on a fine grid of wavenumbers with spacing $\Delta k = 2\pi/2 \tau_0$, which interpolates from solved perturbation modes on a coarse grid $\Delta k = 8/\tau_0$.
Both grids range between $0.1 l_\text{min}/\tau_0 \leq k \leq 3 l_\text{max}/\tau_0$, where $l_\text{min}$ and $l_\text{max}$ are the angular spectrum's minimum and maximum requested multipoles.

\subsection{Adaptive source function refinement}
\label{sec:adaptive}

SymBoltz can optionally refine source functions $S(\tau,k)$ in $k$, such as when computing the matter or CMB power spectra.
First, the source function $S_i(\tau) = S(\tau,k_i)$ is found on an initial grid of wavenumbers $k_i$ by integrating their perturbations.
The method then iterates over each interval $(k_i, k_{i+1})$, integrates the perturbations for the center $k_{i+1/2} = (k_i+k_{i+1})/2$, and finds $S_{i+1/2}(\tau) = S(\tau,k_{i+1/2})$.
The idea is then to compare this to a linearly interpolated source $S_\text{int}(\tau) = (S_i(\tau) + S_{i+1}(\tau)) / 2$ with some error measure $\texttt{err}$, for example from the $L_2$-like $\texttt{err}^2 = \int \mathrm{d}\tau \big( S_\text{int}(\tau) - S_{i+1/2}(\tau) \big)^2$.
If the interpolation is poor and \texttt{err} is greater than some absolute and relative tolerances, the left and right half-intervals $(k_i,k_{i+1/2})$ and $(k_{i+1/2},k_{i+1})$ are recursively refined in the same way.

This gives $S(\tau,k)$ while integrating as few $k$–modes as possible.
Instead of manually creating a nonuniform $k$-grid with many hyperparameters, the process is controlled by only one tolerance.
It parallelizes over each interval and can also be done in $\log k$.

\section{Precision parameters for CLASS}
\label{sec:precision}

When comparing results to CLASS in Sect. \ref{sec:examples}, we use CLASS version 3.3.4 with the following precision parameters:
\begin{codebox}
\begin{Verbatim}
tight_coupling_approximation = 5 # compromise_CLASS; cannot disable TCA
tight_coupling_trigger_tau_c_over_tau_h = 0.01 # turn off TCA earlier
tight_coupling_trigger_tau_c_over_tau_k = 0.001 # turn off TCA earlier
radiation_streaming_approximation = 3 # turn off RSA
ncdm_fluid_approximation = 3 # turn off NCDMFA
ur_fluid_approximation = 3 # turn off UFA

evolver = 1 # for implicit ndf15; 0 for explicit rk
tol_perturbations_integration = 1.0e-6 # less noise; varied in work-precision diagram
tol_ncdm_newtonian = 1.0e-5 # default sampling of massive neutrino momenta
background_Nloga = 6000 # for stable derivatives
\end{Verbatim}
\end{codebox}
We run CLASS with \verb|output = mPk| to compute only $P(k)$, and \verb|output = tCl, pCl| to compute only $C_l^\text{TT}$, $C_l^\text{EE}$ and $C_l^\text{TE}$ with the code's most specialized and optimized paths.
CLASS and SymBoltz are configured to use the same Newtonian gauge, $\lmax$ for all species, RECFAST recombination model, $\tanh$-like reionization parameterization, $w_0 w_a$ dark energy equations, and cosmological parameter values.
The settings above disable as many approximations as possible and reduce the impact of the tight-coupling approximation, which cannot be disabled.
We had to decrease the parameter \verb|background_Nloga| from the default value 40000 to make the derivatives in Fig. \ref{fig:derivatives} stable.
This parameter controls the number of points used for splining background functions in the perturbations.
Its default value was changed from 3000 to 40000 in 2023, but we suspect that the increased density in points makes the splines oscillate from numerical noise.
Likewise, the tolerance for the perturbations ODEs is set one notch smaller than the default $10^{-5}$ to make the output stable.
We use CLASS' default sampling of massive neutrino momenta, as discussed in Sect. \ref{sec:performance}.

When running CLASS with all approximations on in Table \ref{tab:timings} and Fig. \ref{fig:workprec}, the first block of parameters above are not used.
In the work-precision comparison in Fig. \ref{fig:workprec}, we also vary \verb|tol_perturbations_integration| and \verb|evolver| as described in the figure.

\section{Testing and comparison to CLASS}
\label{sec:testing}

SymBoltz' code repository is set up with continuous integration that runs several tests and builds updated documentation pages every time changes to the code are committed.
In particular, this compares the solution for \LCDM{} with CLASS for many variables solved by the background, thermodynamics, and perturbations (using the options \texttt{write\_background}, \texttt{write\_thermodynamics}, and \texttt{k\_output\_values}).
These are the basis for all derived quantities, such as luminosity distances, matter, and CMB power spectra, which are also compared.
The checks pass when the quantities agree within a small tolerance.
We do not compare directly against additional codes such as CAMB, but CLASS has already been compared extensively with CAMB with excellent agreement \citep{lesgourguesCosmicLinearAnisotropy2011c}.
This comparison is found in SymBoltz' documentation.\footnote{\url{https://hersle.github.io/SymBoltz.jl/stable/comparison/}}

Another test checks that integration of the background and perturbations equations is stable throughout parameter space.
As SymBoltz integrates stiff equations without approximations, we could imagine that the integration would be stable for some parameter values and unstable for others.
The test creates a box in parameter space $\pm 50\%$ around a fiducial set of realistic parameter values, draws several sets of parameter values from that space with Latin hypercube sampling (to efficiently cover the box), and integrates the background and perturbations for each such set.
All samples are found to integrate successfully.

\end{document}